\documentclass{aa}

\usepackage{graphicx}
\usepackage{txfonts}
\usepackage{hyperref}

\bibpunct{(}{)}{;}{a}{}{,}
\newcommand{\angstrom}{\textup{\AA}}
\newcommand\HI{H\,\textsc{i}}
\newcommand\HII{H\,\textsc{ii}}

\defcitealias{PlanckCollaboration2014}{Planck Collaboration et al. (2014)}

\begin{document}

\title{From atoms to stars: Modelling $\mathrm{H}_2$ formation \\ and its impact on galactic evolution}

\author{
    E.~Lozano \inst{\ref{inst1}} \and
    C.~Scannapieco \inst{\ref{inst1}} \and
    S.E.~Nuza \inst{\ref{inst2}} \and
    Y.~Ascasibar \inst{\ref{inst3},\ref{inst4}} \and
    V.~Springel \inst{\ref{inst5}}
}

\institute{
    Universidad de Buenos Aires, Facultad de Ciencias Exactas y Naturales, Departamento de Física, Buenos Aires, Argentina \\ \email{elozano@df.uba.ar} \label{inst1} \and
    Instituto de Astronomía y Física del Espacio (IAFE, CONICET-UBA), CC 67, Suc. 28, 1428 Buenos Aires, Argentina\label{inst2} \and
    Departamento de Física Teórica, Universidad Autónoma de Madrid (UAM), Campus de Cantoblanco, E-28049 Madrid, Spain \label{inst3} \and
    Centro de Investigación Avanzada en Física Fundamental (CIAFF-UAM), 28049 Madrid, Spain \label{inst4} \and
    Max Planck Institute for Astrophysics, Karl-Schwarzschild-Str. 1, D-85748 Garching, Germany \label{inst5}
}

\date{Received: 29 March 2025 / Accepted: 05 September 2025}

\abstract{We present a sub-grid model for star formation in galaxy simulations, incorporating molecular hydrogen ($\mathrm{H}_2$) production via dust grain condensation and its destruction through star formation and photodissociation. Implemented within the magnetohydrodynamical code AREPO, our model tracks the non-equilibrium mass fractions of molecular, ionised, and atomic hydrogen, as well as a stellar component, by solving a system of differential equations governing mass exchange between these phases. Star formation is treated with a variable rate dependent on the local $\mathrm{H}_2$ abundance, which itself varies in a complex way with key quantities such as gas density and metallicity. Testing the model in a cosmological simulation of a Milky Way-mass galaxy, we obtain a well-defined spiral structure at $z = 0$, including a gas disc twice the size of the stellar one, alongside a realistic star formation history. Our results show a broad range of star formation efficiencies per free-fall time, from as low as $0.001\%$ at high redshift to values between $0.1\%$ and $10\%$ for ages $\gtrsim 3-4 \, \mathrm{Gyr}$. These findings align well with observational estimates and simulations of a turbulent interstellar medium. Notably, our model reproduces a star formation rate versus molecular hydrogen surface densities relation akin to the molecular Kennicutt-Schmidt law. Furthermore, we find that the star formation efficiency varies with density and metallicity, providing an alternative to fixed-efficiency assumptions and enabling comparisons with more detailed star formation models. Comparing different star formation prescriptions, we find that in models that link star formation to $\mathrm{H}_2$, star formation onset is $\sim \! 500 \, \mathrm{Myr}$ later than those relying solely on total or cold gas density.}

\keywords{galaxies: formation -- galaxies: evolution -- galaxies: ISM -- galaxies: star formation -- methods: numerical}

\maketitle

\section{Introduction}
\label{sec:introduction}

In the context of the standard cosmological model, the formation of galaxies results from the condensation of gas within dark matter haloes, which gravitationally collapse from small density perturbations and later grow through the aggregation of matter and smaller substructures \citep{Rees1977,Silk1977,White1978,Blumenthal1984}. Since the early stages of galaxy formation, several physical processes have been shaping the properties of galactic systems, acting as fundamental components of the intricate network of mechanisms unfolding across cosmic time. Gravitational collapse, gas cooling, mass accretion, star formation (SF), chemical contamination, and various forms of feedback act together, in an interconnected way, determining the properties of galaxies as they evolve \citep{Dalgarno1972,Efstathiou1992,White1991}. The occurrence of mergers, interactions, and mass accretion might also trigger morphological transformations and variations in the dynamical, chemical, and structural properties of galactic systems.

While disentangling the complex interplay of these processes and their individual effects on the evolution of galaxies remains elusive, numerical simulations developed over the last decades have proven instrumental in understanding how galaxies form and evolve within a cosmological context. Current simulations can produce systems that resemble the overall properties of real galaxies, including our Galaxy \citep{Grand2017,Springel2018,Libeskind2020}.

One of the crucial ingredients in simulations of the growth and evolution of galaxies is the proper modelling of SF and the corresponding feedback. Stars form as the result of the collapse and fragmentation of dense gas clouds in the centre of dark matter haloes along the discs of spiral galaxies. Initially triggered by gas cooling -- which determines the amount and distribution of dense gas -- SF works as a source of chemical elements and energy for the interstellar medium (ISM). The balance between cooling and the heating resulting from supernova explosions is key in determining the star formation rates (SFRs) over time and space. Additionally, factors such as the circulation of gas due to feedback, accretion of fresh gas from the intergalactic medium and galactic fountains, internal instabilities, and interactions and collisions influence the ISM properties in non-trivial ways. One of the most important advantages of cosmological simulations is that all these processes participating in the evolution of galaxies are naturally accounted for. However, neither SF nor stellar feedback can be included from first principles, as the scales at which they occur are not resolved. For this reason, sub-grid recipes are included to account for the impact of unresolved physics on larger scales, thereby allowing galaxy formation simulations to be run on cosmological volumes at high resolutions.

In the case of SF, models assuming a simple correlation between the SFR density and the gas density have been used since the pioneering studies of \cite{Katz1992} and \cite{Steinmetz1994}. In such models, cold dense gas transforms into star particles at a rate that is proportional to the gas density and inversely proportional to a typical timescale. This approach, if accompanied by an efficient modelling of feedback, predicts galaxies that reproduce the slope of the observed relation between the SFR and the gas density \citep{Schmidt1959,Kennicutt1998}, as well as the SFR levels, as long as a tuneable proportionality factor is introduced, known as the star formation efficiency (SFE), and the typical SF timescale is the dynamical or free-fall time of the gas \citep{Scannapieco2006}.

During the last decade, more recent observations made it evident that the correlation between the gas density and the SFR is even stronger when molecular gas is considered (\citealt{Wong2002,Leroy2008,Bolatto2011,Schruba2011} and references therein), both at resolved scales \citep{Baker2021} and at integrated (i.e. galaxy-wide) scales across different redshifts \citep{Baker2022}. Additionally, the typical timescale of SF may vary depending on the properties of the ISM where SF occurs \citep{Bigiel2008,Bigiel2010,Bolatto2011,Leroy2013}. Motivated by these results, a number of studies have developed more sophisticated SF models within simulations and semi-analytic models. These models rely on a better characterisation of the ISM (either increasing numerical resolution or developing better sub-grid models), opening up the possibility to follow the content of molecular hydrogen as a function of time and to introduce $\mathrm{H}_2$-based SF laws \citep{Murante2010,Molla2015,Millan-Irigoyen2020}.

In contrast to SF, modelling the effects of feedback in simulations has proven extremely complex. In part, the reason for this is that a detailed understanding of stellar feedback and its interaction with the surrounding medium has not been fully achieved, and discussion on numerical aspects of the modelling still remain (see e.g. \citealt{Agertz2013}). Moreover, different ways of implementing stellar feedback show significant variations in predictions for the stellar masses, morphologies, and other galaxy properties, even for a given initial condition (IC) \citep{Scannapieco2012}, although significant progress has been made in this direction \citep{Kim2013,Kim2016}. The modelling of feedback is naturally linked to the modelling of SF, and thus the various numerical parameters that are included to regulate SF and feedback levels are unique to a given model, lacking any corresponding physical parameter whose value can be estimated in experiments or observations. Finally, additional forms of feedback including black holes and cosmic rays have also been explored, as they also participate in the process of galaxy formation but may not be as important as stellar feedback for galaxies up to the Milky Way (MW) mass scale \citep{Naab2017}.

While feedback is what ultimately regulates the SF activity in galaxies, it is clear that an adequate description of the ISM is key to determine the conditions for SF. Various sub-resolution models, which describe the multi-phase structure of the ISM, have been proposed, with the aim of providing a better modelling of SF in the context of simulations. \cite{Yepes1997} and \cite{Hultman1999} incorporated in Eulerian and Lagrangian codes the \cite{McKee1977} model, assuming that each gas element was composed of a hot and cold phase in pressure equilibrium. Cold clouds form in the hot medium due to thermal instability and produce stars, which transfer mass back to the hot phase and can evaporate the cold clouds. \cite{Springel2003} improved and extended these models in the \texttt{GADGET} code, also adding galactic winds driven by SF as a form of feedback.

As for numerical implementations including molecular hydrogen, in \cite{Pelupessy2006} a sub-grid model for the formation of molecular gas was implemented, using a time-dependent model for the \HI $\,\, \rightarrow \mathrm{H}_2$ transition. The model was used for an integrated study of SF and $\mathrm{H}_2$ gas on a dwarf galaxy, in the setting of non-cosmological galaxy models. In the isolated galaxy formation simulations of \cite{Robertson2008}, $\mathrm{H}_2$ was tracked assuming equilibrium abundances based on the gas metallicity. \cite{Gnedin2009} implemented a model for tracking the non-equilibrium abundances of $\mathrm{H}_2$ in their adaptive mesh refinement (AMR) code and ran cosmological simulations of low-metallicity galaxies, exploring the interactions between atomic gas, molecular gas, and dust. \cite{Christensen2012} adapted this model to incorporate it into their smoothed particle hydrodynamics (SPH) code and ran simulations of isolated dwarf galaxies in a cosmological context up to $z = 0$. \cite{Murante2014} implemented a molecular-based SF law in cosmological simulations of a MW-type halo. In their model, SF is linked to the molecular abundance within gas particles computed using the phenomenological relation of \cite{Blitz2006}, which only depends on gas pressure. \cite{Valentini2022} used cosmological simulations of a MW-size halo, focusing on the low-metallicity regime, to compare the predictions of two different models for computing the molecular fraction: the theoretical model of \cite{Krumholz2009} and the phenomenological approach of \cite{Blitz2006}. They found that the former model produces a clumpier ISM and a more complex $\mathrm{H}_2$ distribution in better agreement with observations of nearby disc galaxies. \cite{Hopkins2014} also assumed an $\mathrm{H}_2$-regulated SF prescription for their galaxy formation simulation, computing the molecular hydrogen density with the local column density and metallicity as in \cite{Krumholz2011}.

In this work, we further explore the influence of assuming a molecular-based SF law on the formation of galaxies. We base our work on the semi-analytical model of \citet{Millan-Irigoyen2020}, which assumes that the ISM has a multi-phase structure that includes atomic, ionised, and molecular gas, as well as metals and dust (which is especially important for molecular hydrogen formation in low metallicity environments at high redshift). Our model is grafted onto the moving-mesh, magnetohydrodynamical code \texttt{AREPO} \citep{Springel2010}, which discretises the gas component into cells. We assume that each of these cells represents a multi-phase structure with ionised, atomic, and molecular hydrogen phases, as well as a stellar phase (note that metals are followed as well, using the standard chemical evolution model of \texttt{AREPO}). The evolution of each of these phases is governed by a set of differential equations, which quantify the amount of mass that is exchanged between them because of different physical mechanisms. The SFR associated with the stellar phase is linked to the fraction of molecular hydrogen and is regulated by a density-dependent SF timescale.

While our implementation of SF is explicitly tied to the molecular hydrogen fraction, we acknowledge that the causal role of molecular gas in SF remains debated. Numerical simulations by \citet{Glover2012} have shown that SF can proceed efficiently in cold atomic gas, and that a cloud's ability to shield itself from the interstellar radiation field -- rather than its chemical composition -- is key to determine whether it can cool and collapse to form stars. Observationally, SF has also been detected in \HI-dominated environments, including dwarf irregular galaxies with little or no detectable molecular gas \citep{Roychowdhury2015}. These findings suggest that $\mathrm{H}_2$ may function more as a tracer of cold, dense gas than as a strict prerequisite for SF. Therefore, the assumption of an $\mathrm{H}_2$-based SF prescription in our model should be interpreted with the caveat that it may not be universally valid, particularly in low-metallicity or poorly shielded environments.

This work is organised as follows. A description of the simulation code, the included physics, and the new SF model is provided in Section~\ref{sec:simulations}. In Section~\ref{sec:results}, we apply our model to the formation of a MW-mass galaxy in a cosmological context and discuss our predictions for the properties of the simulated system, focusing on the molecular, atomic, and ionised phases, and on the SF activity. In Section~\ref{sec:ks_law_epsilon_ff} we present the predictions of our model for the molecular Kennicutt-Schmidt (KS) law and for the SFE per free-fall time associated with gas cells, and in Section~\ref{sec:discussion} we discuss the implications of assuming a molecular-based SF law. Finally, in Section~\ref{sec:conclusions}, we give our conclusions.

\section{The simulation code and the star formation model}
\label{sec:simulations}

In this section, we describe the implementation of our SF sub-grid model, which follows the ionised, atomic, and molecular phases of hydrogen and links the SFR to the molecular phase. We first summarise the main characteristics of the base code where we implemented our model, and then, the model itself.

\subsection{The \texttt{AREPO} code}

We used the $N$-body, magnetohydrodynamics (MHD) code \texttt{AREPO} to perform the simulations, which we briefly describe here. For a more comprehensive explanation, we refer to \cite{Springel2010}. \texttt{AREPO} is a moving-mesh code that tracks MHD and collisionless dynamics in a cosmological setting. Gravitational forces are computed using a conventional TreePM method \citep{Springel2005}, which employs a Fast Fourier Transform technique for long-range forces and a hierarchical oct-tree algorithm \citep{Barnes1986} for short-range forces, in conjunction with adaptive time-stepping. \texttt{AREPO} employs a dynamic unstructured mesh, constructed from a Voronoi tessellation of the simulation box. This allows for a finite-volume discretisation of the MHD equations. The MHD equations are solved using a second-order Runge–Kutta integration scheme with high-accuracy least-squares spatial gradient estimators of primitive variables \citep{Pakmor2015}. 

A distinctive feature of \texttt{AREPO} is the ability to transform the mesh at any time step through a mesh reconstruction, a capability not found in standard grid-based methods. The mesh reconstruction ensures that each cell contains a specific target mass (within a certain tolerance), meaning that areas of high density are resolved with more cells than areas of low density. Additionally, the mesh generating points move with the fluid flow, allowing \texttt{AREPO} to overcome the Galilean non-invariance problem that standard Eulerian mesh codes have and significantly reduce the advection errors that appear in complex supersonic flows. The quasi-Lagrangian nature of the method makes it comparable to other Lagrangian methods such as SPH, although it eliminates several of its limitations, such as the need for artificial viscosity.

\texttt{AREPO} provides several modules to model different physical phenomena. In particular, we apply the galaxy formation model used in the Auriga Project, which includes primordial and metal-line cooling, a uniform ultraviolet background field for reionisation, magnetic fields \citep{Pakmor2014,Pakmor2017,Pakmor2018}, energetic and chemical feedback from Type II supernovae, and mass loss and metal return owing to Type Ia supernovae and asymptotic giant branch stars \citep{Vogelsberger2013,Marinacci2013,Grand2017}. We did not include the active galactic nuclei feedback module, given that its effects are subdominant for a MW-mass galaxy.

Our code, including the SF, feedback and cooling routines, is identical to the standard \texttt{AREPO} implementation, except for the calculation of the SF probability associated with each gas cell (see next section). Using this probability, star particles are formed stochastically from gas cells that are eligible to do so. As we only vary the SF probability leaving the rest of the implementation unchanged, any difference in the results when our code is used will only be due to the changes induced in the SF and subsequent feedback levels, and their effects on the spatial distribution and properties of the ISM.

\subsection{The star formation model}
\label{subsec:model_equations}

In our SF model, star particles are formed from gas cells following a stochastic approach. In this scheme, described in \cite{Springel2003} (see also \citealt{Grand2017}), gas elements that are denser than a density threshold ($\rho_\mathrm{th}$) are eligible to form stars, with a probability given by
\begin{equation}
    \label{eq:SF_probability}
    p = 1 - \exp\left(- \frac{\dot{m}_* \, \Delta t}{M_\mathrm{cell}}\right) \, ,
\end{equation}
where $\dot{m}_*$ is the SFR associated with the gas cell of mass $M_\mathrm{cell}$ and $\Delta t$ is the integration time step. The exponent in Eq. (\ref{eq:SF_probability}), $\dot{m}_* \, \Delta t \, / M_\mathrm{cell}$, thus represents the fraction of stellar mass formed in the cell during the time step.

In the standard formulation of \texttt{AREPO}, the SFR of a gas cell with density $\rho$ and cold mass fraction $x$ is given by
\begin{equation} 
    \dot{m}_* \biggr\rvert_\mathrm{STD} = \frac{x \, M_\mathrm{cell}}{t_*(\rho_\mathrm{cell})} \, ,
\end{equation}
where $t_*$, the SF timescale, is
\begin{equation}
    t_*(\rho_\mathrm{cell}) = t_0^* \left(\frac{\rho_\mathrm{cell}}{\rho_\mathrm{th}}\right)^{\!-1/2} \, ,
\end{equation}
and $t_0^*$, the maximum SF timescale, is a parameter of the model.

In our model, instead, the probability of forming a star is
\begin{equation} 
    p = 1 - \exp(-f_s) \, ,
\end{equation}
where $f_s$ is the fraction of stellar mass in a gas cell. This fraction is obtained by solving a system of differential equations, which follows four components: an ionised phase with a temperature of $\sim \! 10^4 \, \mathrm{K}$, an atomic phase with a temperature of $\sim \! 100 \, \mathrm{K}$, a molecular component with a temperature of $\sim \! 10 \, \mathrm{K}$, and stars\footnote{In the model, we only consider the dominant channel of the reactions that exchange mass between the different phases. For this reason, even though the gas is made up of hydrogen, helium, and metals, only hydrogen reactions are incorporated into the model.}. The four components are parametrised as mass fractions relative to the total cell mass: $f_X = M_X / M_\mathrm{cell}$ where $X$ can be $i$: ionised, $a$: atomic, $m$: molecular, or $s$: stellar. The mass of the cell at each time $t$ is therefore
\begin{equation}
    M_\mathrm{cell}(t) = M_i(t) + M_a(t) + M_m(t) + M_s(t) \, ,
\end{equation}
and the sum of the corresponding fractions add up to unity
\begin{equation}
    \label{eq:fractions}
    f_i + f_a + f_m + f_s = 1 \, .
\end{equation}

\begin{figure}[ht]
    \includegraphics[width=8.8cm]{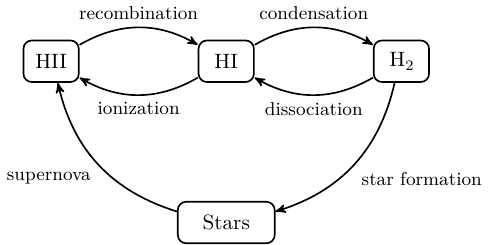}
    \caption{Physical processes responsible for transformations between the different phases. Arrows going inwards indicate that the named process increases the amount of the corresponding phase, and vice versa for arrows going outward.}
    \label{fig:physical_processes}
\end{figure}

It is important to emphasise that these fractions represent the internal composition of each gas cell within our sub-grid model, rather than physical particle masses in the simulation. In particular, the stellar mass fraction ($f_s$) does not correspond to actual star particles; instead, it is used solely to compute the instantaneous SFR of a gas cell, which in turn governs whether a new star particle is created. Moreover, in practice we find that $f_a + f_m + f_i = 1 - f_s \approx 1$, since $f_s$ remains small in all gas cells, typically of the order of $3 \times 10^{-4}$. As illustrated in Fig.~\ref{fig:physical_processes}, the four phases exchange mass via the following physical processes: recombination of ionised atoms, condensation of atomic hydrogen, photodissociation of molecular gas, photoionisation of atoms, SF, and mass return from stars to the interstellar gas.

In terms of the mass fractions, the equations governing the evolution of the phases are
\begin{subequations}
    \label{eq:odes}
    \begin{align}
        \frac{\mathrm{d}}{\mathrm{d}t}f_i(t) &= - \frac{f_i(t)}{\tau_\mathrm{rec}(t)} + \eta_\mathrm{ion} \, \psi(t) + R \, \psi(t) \, , \\
        \frac{\mathrm{d}}{\mathrm{d}t}f_a(t) &= \frac{f_i(t)}{\tau_\mathrm{rec}(t)} - \eta_\mathrm{ion} \, \psi(t) - \frac{f_a(t)}{\tau_\mathrm{cond}(t)} + \eta_\mathrm{diss} \, \psi(t) \, , \\
        \frac{\mathrm{d}}{\mathrm{d}t}f_m(t) &= \frac{f_a(t)}{\tau_\mathrm{cond}(t)} - \eta_\mathrm{diss} \, \psi(t) - \psi(t) \, , \\
        \frac{\mathrm{d}}{\mathrm{d}t}f_s(t) &= \psi(t) - R \, \psi(t) \, ,
    \end{align}
\end{subequations}
where $\tau_\mathrm{rec}$ and $\tau_\mathrm{cond}$ are the time-scales for recombination and condensation, respectively, $\eta_\mathrm{ion}$ and $\eta_\mathrm{diss}$ represent the efficiencies of ionisation and dissociation of hydrogen, respectively, $R$ is the mass return fraction from stars, and $\psi(t)$ is the fractional instantaneous SFR, 

\begin{equation}
    \label{eq:sfr}
    \frac{\mathrm{d}}{\mathrm{d}t} f_s(t) \biggr\rvert_{\mathrm{SFR}} =: \psi(t) = \frac{f_m}{\tau_\mathrm{star}} \, .
\end{equation}

We assume it depends only on the amount of molecular gas and is regulated by a SF timescale ($\tau_\mathrm{star}$), as in \citet{Millan-Irigoyen2020}.

The exchange of mass between phases due to recombination, condensation, and SF are modelled using timescales ($\tau_\mathrm{rec}$, $\tau_\mathrm{cond}$, and $\tau_\mathrm{star}$, respectively), which depend on the properties of the gas and are therefore time- and cell-dependent. In contrast, the terms corresponding to photoionisation, photodissociation, and mass return are proportional to $\psi(t)$, with constant proportionality factors ($\eta_\mathrm{ion}$, $\eta_\mathrm{diss}$, and $R$, respectively). A detailed description and computation of all these factors is provided in Section~\ref{subsec:parameters}.

Our system of equations was solved for each gas cell that is eligible to form stars, and integrated during the corresponding time step, with the assumption that the mass, density, and metallicity of each gas cell was constant during integration. Gas cells enter our routine with initial values for the ionised and atomic fractions given by the standard treatment of ionised and neutral phases in \texttt{AREPO}, provided that they have neither molecular nor stellar mass. It is worth noting that because \texttt{AREPO} is not purely Lagrangian, there might be mass fluxes between cells. If inflowing gas into a cell does not have the same relative mass fractions than the present cell, the $f_X$ values should in principle be affected. For now, we ignored such fluxes and treated every cell as a 'closed box' during the integration of our set of equations.

It is important to note that the density threshold for SF, $\rho_\mathrm{th}$, which determines whether a gas cell enters our SF routine at each time step, does not have a large impact on our results. While we adopted a fiducial value of $0.19 \, \mathrm{cm^{-3}}$, as used in cosmological simulations of similar resolution, gas cells with densities below $\rho_\mathrm{th}$ would have a very low stellar fractions (as we show in appendix~\ref{app:fixed_ics}) and therefore a very low probability of producing star particles. We tested this with a simulation using $0.019 \, \mathrm{cm^{-3}}$ for the $\rho_\mathrm{th}$, and getting very similar results compare to our fiducial simulation. In other words, the minimum density of a gas cell needed to produce stars is encoded in our SF law via the link between SF and the amount of molecular gas in the cell.

\subsection{Parameters of the model}
\label{subsec:parameters}

\subsubsection{Star formation timescale}

For the SF timescale we use the free-fall time of the gas cell, as in \cite{Krumholz2005}
\begin{equation}
    \tau_\mathrm{star} = \sqrt{\frac{3\pi}{32 \, G \, \rho_\mathrm{g}}} \, ,
\end{equation}
where $G$ is the gravitational constant and $\rho_\mathrm{g}$, in the context of our model, is the density of the corresponding cell, $\rho_\mathrm{cell}$. It is worth noting that we adopt the free-fall time of the gas cell as the star formation timescale, following the definition in \cite{Krumholz2005}. However, unlike their formulation, we do not include a separate efficiency parameter $\epsilon_\mathrm{ff}$; instead, SF is self-regulated in our model via the coupled evolution of the gas phases described by Eqs.\ (\ref{eq:odes}), which naturally determines the effective SFR.

\subsubsection{Recombination timescale}

The recombination rate of hydrogen atoms can be written as \citep{Osterbrock2006,Gnedin2009,Christensen2012,Millan-Irigoyen2020}
\begin{equation}
    \frac{\mathrm{d}}{\mathrm{d}t} n_a \biggr\rvert_\mathrm{rec} = \frac{n_\mathrm{i}}{\tau_\mathrm{rec}} = \langle \sigma \, v \rangle_\mathrm{B} \, n_e \, n_i \, ,
\end{equation}
where $\tau_\mathrm{rec}$ is the recombination timescale and $\langle \sigma \, v \rangle_\mathrm{B}$ is the thermally averaged Case-B recombination cross-section (we do not consider transitions to the ground state). We take a value of $\langle \sigma \, v \rangle_\mathrm{B} = 2.6 \times 10^{-13} \, \mathrm{cm^3 \, s^{-1}}$, which is appropriate for a ionised phase with a temperature of $10^4 \, \mathrm{K}$ \citep{Nebrin2023}.
$n_e$ is the electron number density, which, by mass balance and within our model, is equal to the ionised hydrogen number density $n_i$. The number densities are calculated from the fractions as
\begin{equation}
    n_X = \frac{N_X}{V_\mathrm{cell}} = \frac{f_X}{m_X} \, \rho_\mathrm{cell} \, ,
\end{equation}
where $\rho_\mathrm{cell}$ is the density of the gas cell, $N_X$ is the number of particles of component $X$, and $m_X$ is the mass of a single element of the $X$ component.

In terms of the fraction of atomic hydrogen, $f_a$, the rate at which atomic hydrogen is created can be written as
\begin{equation}
    \frac{\mathrm{d}}{\mathrm{d}t} f_\mathrm{a} \biggr\rvert_\mathrm{rec} =
    \langle \sigma \, v \rangle_\mathrm{B} \, \frac{\rho_\mathrm{cell}}{m_P} \, f_i^{\,2} = \frac{f_i}{\tau_\mathrm{rec}} \, ,
\end{equation}
where $m_P$ is the proton mass and we defined the timescale $\tau_\mathrm{rec}$ as
\begin{equation}
    \tau_\mathrm{rec} = \frac{m_P}{\langle \sigma \, v \rangle_\mathrm{B} \, \rho_\mathrm{cell} \, f_i} \, .
\end{equation}

It is important to note that this is not a constant, as it depends not only on the density of the cell, but also on the ionised fraction $f_i$.

\subsubsection{Condensation timescale}

The rate of molecular hydrogen formation due to the condensation of atomic gas in the surface of dust grains can be written as \citep{Hollenbach1971a,Hollenbach1971b}
\begin{equation}
    \label{eq:dt-n_m}
    \frac{\mathrm{d}}{\mathrm{d}t} n_m \biggr\rvert_\mathrm{cond} = R_d \, n_H \, n_a \, ,
\end{equation}
where $R_d$ is the dust grain $\mathrm{H}_2$ formation rate coefficient, $n_H$ is the hydrogen number density, and $n_a$ is the atomic hydrogen number density.

In the literature $n_H$ is generally taken as $n_H = n_a + 2 \, n_m$, because most studies consider cold gas clouds ($T \sim 100 \, \mathrm{K}$) dominated by molecular and atomic hydrogen. In contrast, we have atomic, molecular, and ionised gas within our gas cell; so, we use $n_H = n_a + 2\, n_m + n_i$.

In terms of our adopted notation, we can write Eq. (\ref{eq:dt-n_m}) as
\begin{equation}
    \frac{\mathrm{d}}{\mathrm{d}t} f_m \biggr\rvert_\mathrm{cond} = 2 \, R_d \, ( f_a + f_m + f_i) \, \frac{\rho_\mathrm{cell}}{m_P} \, f_a \, .
\end{equation}

Similarly to our approach of the previous section, we can rewrite this equation as
\begin{align}
    \frac{\mathrm{d}}{\mathrm{d}t} f_m \biggr\rvert_\mathrm{cond} = \frac{f_a}{\tau_\mathrm{cond}} \, ,
\end{align}
and define a condensation timescale, $\tau_\mathrm{cond}$, as
\begin{equation}
    \tau_\mathrm{cond} = \frac{m_P}{2 \, R_d \, \rho_\mathrm{cell} \, (f_a + f_m + f_i)} \, ,
\end{equation}
which depends on the fractions of atomic, molecular, and ionised hydrogen.

Theoretical and experimental work has shown that $R_d \propto T^{1/2}$ \citep{Black1987}, and $R_d \propto n_\mathrm{dust}$, the number density of dust grains \citep{Hollenbach1971b}. Assuming the simplest dust model, where the amount of dust is proportional to the metallicity $Z$, we have $R_d \propto T^{1/2} \, Z$ \citep{Pelupessy2006,Wolfire2008}. However, following previous prescriptions \citep{Pelupessy2006,Gnedin2009,Christensen2012,Molla2017,Millan-Irigoyen2020}, we assume that $R_d$ only scales with the metallicity, using $\sim \! 100 \, \mathrm{K}$ for the temperature of the cold neutral gas. We also normalise $R_d$ to the solar metallicity \citep{Wolfire2008}
\begin{equation}
    R_\odot = 3.5 \times 10^{-17} \, \mathrm{cm^3 \, s^{-1}} \, ,
\end{equation}
and thus we have
\begin{equation}
    R_d = \frac{R_{\odot}}{Z_\odot} \, Z \, ,
\end{equation}
where $Z_\odot = 0.0127$, the same value used within \texttt{AREPO}.

It is important to note that, in the case of uncontaminated gas ($Z = 0$), $R_d$ goes to $0$ and thus, $\tau_\mathrm{cond}$ diverges. This is solved by introducing an effective metallicity $Z_\mathrm{eff}$, which is added to $Z$, resulting in the following expression for $\tau_\mathrm{cond}$
\begin{equation}
    \label{eq:tau_cond}
    \tau_\mathrm{cond} = \frac{m_P \, Z_\odot}{2 \, R_\odot \, (Z + Z_\mathrm{eff}) \, \rho_\mathrm{cell} \, (f_a + f_m + f_i)} \, .
\end{equation}

Adding an effective metallicity has also a physical meaning. If $R_d \propto Z$, the underlying hypothesis is that the only conversion channel for \HI $\,\, \rightarrow \mathrm{H}_2$ is the condensation on the surface of dust grains. Assuming instead $R_d \propto {Z + Z_\mathrm{eff}}$, we also consider molecular formation at very low metallicities. Following \cite{Glover2007}, we use $Z_\mathrm{eff} = 10^{-3} \, Z_\odot$.

Finally, we use an additional adjustable parameter, the clumping factor $C_\rho = \langle \rho^2 \rangle / \langle \rho \rangle^2$ \citep{Gnedin2009,Christensen2012} with a fiducial value of $C_\rho = 100$, in order to phenomenologically account for any potential uncertainties in $\tau_\mathrm{cond}$. So, we end up with
\begin{equation}
    \tau_\mathrm{cond} = \frac{m_P \, Z_\odot}{2 \, R_\odot \, C_\rho \, (Z + Z_\mathrm{eff}) \, \rho_\mathrm{cell} \, (f_a + f_m + f_i)} \, .
\end{equation}

It is worth noting again that $\tau_\mathrm{cond}$ is not a constant value, as depends not only on the density, but on the atomic, molecular, and ionised fractions (or, equivalently, on the stellar fraction, as these four fractions add up to unity) and on the metallicity.

\subsubsection{Photodissociation and photoionisation efficiencies}

The disassociated and ionised mass rates per unit of created stellar mass are
\begin{align}
    \eta_\mathrm{diss} & = \frac{\dot{M}_\mathrm{diss}}{\mathrm{SFR}} = 2 \, m_P \, c_\mathrm{diss} \, \frac{\dot{N}_\mathrm{diss}}{\mathrm{SFR}} \, \, , \\
    \eta_\mathrm{ion}  & = \frac{\dot{M}_\mathrm{ion}}{\mathrm{SFR}} = m_P \, \frac{\dot{N}_\mathrm{ion}}{\mathrm{SFR}} \, ,
\end{align}
where $\dot{N}_\mathrm{diss}$ is the number of photodissociating photons produced per unit time (in the Lyman–Werner band, $912 \, \angstrom$ to $1107 \, \angstrom$), $\dot{N}_\mathrm{ion}$ the number of ionising photons produced per unit time (between $0$ and $912 \, \angstrom$), and $c_\mathrm{diss}$ is an efficiency factor. For the molecular dissociation reaction, we consider the efficiency given by \cite{Draine1996}, where it is shown that dust grains may absorb up to $\sim \! 60$ percent of the photons capable of dissociating hydrogen molecules and a large fraction of the remaining photons excite different rotational and vibrational states, reducing their dissociation probability to $\sim \! 15$ percent. We thus have $c_\mathrm{diss} = 0.4 \times 0.15 = 0.06$. For the ionisation reaction, we assume $100 \%$ efficiency.

Using the number of photons produced per unit time, $Q(t', Z')$, by a single stellar population of $1 \, \mathrm{M_\odot}$, age $t'$, and metallicity $Z'$, we can calculate $\dot{N}$ as
\begin{equation}
    \dot{N}(t) = \int_0^t \mathrm{SFR}(t - t') \, Q(t', Z') \, \mathrm{d}t' \, ,
\end{equation}
where $\mathrm{SFR}(t - t')$ is the instantaneous SFR at the moment of birth of the stellar population of age $t'$, and $Z' = Z(t - t')$ is defined as the metallicity at that moment. Notice how the integral is in stellar age $t'$, from 0 to the present time $t$. Because most of the contributions to the integral comes from young blue stars that die in the first $10 \ \mathrm{to} \ 100 \, \mathrm{Myr}$ ($Q$ is up to 6 orders of magnitude larger for young stars than for older ones) it is possible to do the approximations
\begin{align}
    \mathrm{SFR}(t - t') &\approx \mathrm{SFR}(t) \, , \\
    Z(t - t') &\approx Z(t) \, ,
\end{align}
if one assumes that the $\mathrm{SFR}$ and $Z$ vary slowly in those timescales.

Then, the number of photons per unit mass and unit time can be written as
\begin{equation}
    Q(t', Z) = \int_{\lambda_1}^{\lambda_2} \frac{\lambda \, L_\lambda(t', Z)}{h \, c} \, \mathrm{d}\lambda \, ,
\end{equation}
where $L_\lambda(t', Z)$ is the luminosity per unit of wavelength of a stellar population of $1 \, \mathrm{M_\odot}$, age $t'$, and metallicity $Z$, and therefore we have
\begin{align}
    \eta_\mathrm{diss} &= 2 \, m_\mathrm{P} \, c_\mathrm{diss} \, \int_0^t Q_\mathrm{LW}(t', Z) \, \mathrm{d}t' \, , \\
    \eta_\mathrm{ion} &= m_\mathrm{P} \, \int_0^t Q_\mathrm{H}(t', Z) \, \mathrm{d}t' \, ,
\end{align}
where the subscript $\mathrm{LW}$ indicates that we consider the Lyman-Werner band for dissociation ($\lambda_1 = 921 \, \angstrom$ and $\lambda_2 = 1107 \, \angstrom$) and the subscript $\mathrm{H}$ indicates that the integration used for photoionisation is between $\lambda_1 = 0$ and $\lambda_2 = 912 \, \angstrom$.

The photoionisation and photodissociation efficiencies ($\eta_{\mathrm{ion}}$, $\eta_{\mathrm{diss}}$) are evaluated locally within each gas cell as a function of its metallicity, rather than being derived from the star particles that may eventually form stochastically. Since our method does not include explicit radiative transfer, it does not capture photon propagation or escape, and should therefore be interpreted with this limitation in mind. Although the underlying physical processes are the ionisation and dissociation of hydrogen, the photon production rate is tied to recent SF. Accordingly, our ionisation rates are expressed per unit SFR and must be multiplied by $\psi(t)$.

Using the values from PopStar \citep{Molla2009}, we compute $Q$ for the \citet{Chabrier2003} IMF at the six metallicities for which we have yield tables -- 0.0001, 0.0004, 0.004, 0.008, 0.02, and 0.05 -- and for ages between $0.1\,\mathrm{Myr}$ and $15\,\mathrm{Gyr}$. These values are used as interpolation tables to compute $\eta_\mathrm{diss}$ and $\eta_\mathrm{ion}$ at a given metallicity and integration time, which we take as constants during the integration of Eqs.\ (\ref{eq:odes}). The asymptotic value of $\eta_\mathrm{diss}$ ranges from $1353$ for the lowest metallicity to $816$ for the highest one, and $\eta_\mathrm{ion}$ ranges from $8489$ for the lowest metallicity to $2360$ for the highest one.

We note that our formulation of the dissociation and ionisation processes implicitly assumes that every photon available contributes directly to altering the local chemical state of the gas, i.e. that the local reservoir of molecular and atomic gas can absorb the full photon budget. This is clearly an approximation: in reality, not all photons result in ionisations or dissociations, and some may escape or be absorbed by dust. In fact, the constant efficiency factor $c_{\mathrm{diss}} = 0.06$ defined above partially accounts for this effect, as it represents the attenuation due to dust shielding. We verified that our assumption is valid in most relevant cases: in approximately $90\%$ of the star-forming gas cells, the mass of atomic gas exceeds the amount required to be ionised given the computed value of $\eta_{\mathrm{ion}}$. Thus, while the current formulation may overestimate feedback coupling in certain edge cases, it remains a reasonable approximation for the bulk of the star-forming regions in our simulations.

\subsubsection{Mass recycling}

We parametrise the mass returned from the stellar to the gas component using the parameter $R$, defined as the mass fraction of a stellar population that is returned to the ISM under the instantaneous recycling approximation (IRA), first explicitly used by \cite{Schmidt1963}. Note that this hypothesis can be relaxed considering the lifetimes of stars, as done in \citep{Millan-Irigoyen2020}, which would make $R$ time-dependent. In our simulations, this would produce an additional computational cost, which is not justified, given the short integration timescales over which we solve our system of differential equations.

The returned mass fraction $R$ of a stellar population with masses between $m + \mathrm{d}m$ and Initial Mass Function (IMF) $\phi(m)$ is \citep{Pagel2009,Matteucci2003}
\begin{equation}
    R = \dfrac{\int_{m_\mathrm{ir}}^{m_\mathrm{max}} (m - m_\mathrm{rem}(m)) \, \phi(m) \, \mathrm{d}m}{\int_{m_\mathrm{min}}^{m_\mathrm{max}} m \, \phi(m) \, \mathrm{d}m} \, ,
\end{equation}
where $m_\mathrm{min}$ and $m_\mathrm{max}$ are the minimum and maximum masses of the IMF, $m_\mathrm{ir}$ is the mass limit for the IRA, and $m_\mathrm{rem}(m)$ is the remnant stellar mass given by the yield model. We use the stellar yields model of \citet{Portinari1998} compiled by \citet{Molla2015}, where the parameters have the following values: $m_\mathrm{ir} = 8 \, M_\odot$, $m_\mathrm{min} = 0.08 \, M_\odot$, and $m_\mathrm{max} = 100 \, M_\odot$. We calculated $R$ for the six metallicities for which we have yield tables (see previous section). $R$ ranges from $0.16$ for the lowest metallicity to $0.18$ for the highest one. We use these values as a interpolation table to compute $R$,, which remain constant during the integration of Eqs.\ (\ref{eq:odes}).

\subsubsection{A comparison of the model timescales regulating mass exchange}
\label{subsec:implementation}

\begin{figure}[ht]
    \centering
    \includegraphics[width=8.8cm]{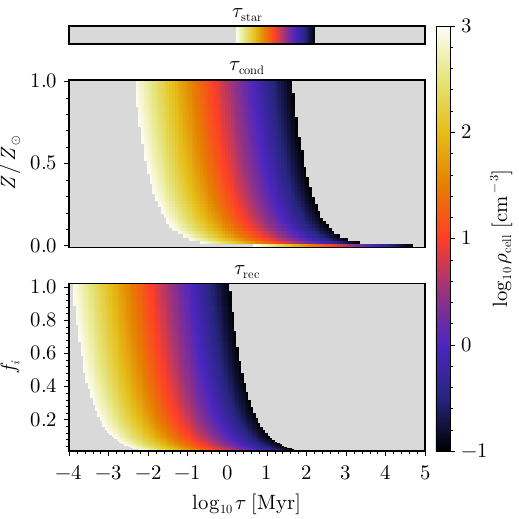}
    \caption{Comparison of the different timescales ($\tau_\mathrm{star}$, $\tau_\mathrm{cond}$, and $\tau_\mathrm{rec}$) in our model. The colour scale indicates the cell density, and the y-axis the secondary dependency of each timescale when appropriate. $\tau_\mathrm{cond}$ depends on the sum $f_a + f_m + f_i$ too (see Section~\ref{subsec:parameters}), but for this plot we take $f_a + f_m + f_i = 1.0$.}
    \label{fig:timescale_comparison}
\end{figure}

The amount of mass exchanged between the different phases, which ultimately determines the SF levels, is regulated by the various timescales involved in the Eqs.\ (\ref{eq:odes}) and their dependencies on the properties of the gas. Fig.~\ref{fig:timescale_comparison} compares the range of possible values for the timescales linked to SF ($\tau_\mathrm{star}$), condensation ($\tau_\mathrm{cond}$), and recombination ($\tau_\mathrm{rec}$). While all three decrease with increasing gas density (the colour scale), $\tau_\mathrm{cond}$ has a secondary dependency on the gas metallicity, and $\tau_\mathrm{rec}$ with the ionised fraction (note that this makes the set of Eqs.\ (\ref{eq:odes}) non-linear). As can be seen from Fig.~\ref{fig:timescale_comparison}, at a given density, $\tau_\mathrm{star}$ is the largest of the three, $\tau_\mathrm{rec}$ the smallest, and $\tau_\mathrm{cond}$ lies in between. Even if we consider the dependencies other than the density, the conversion from ionised to the atomic material is almost always the fastest, followed by the transition from atomic to molecular gas, and finally from the molecular to the stellar phase. This behaviour is what determines the predictions of the model on individual gas cells, as we discussed in more detail in appendix~\ref{app:fixed_ics}.

\subsection{Simulation setup and initial conditions}

We use for our simulations the ICs of the halo denoted as Au6 in the Auriga Project \citep{Grand2017}. These correspond to zoom-in ICs of a halo that has, at $z = 0$, a virial mass of $M_{200} = 1.05 \times 10^{12} \, \mathrm{M_\odot}$ and is relatively isolated. The mass resolution of the simulation (level 4 in Auriga) is $\sim \! 3 \times 10^5 \, \mathrm{M_\odot}$ for dark matter particles and  $\sim \! 5 \times 10^4 \, \mathrm{M_\odot}$ for baryons. 

The gravitational softening length for dark matter and star particles is fixed in comoving coordinates at $500 \, h^{-1} \, \mathrm{pc}$ up to $z = 1$, and fixed in physical units at $369 \, \mathrm{pc}$ thereafter. The softening length of gas cells scales with the mean radius of the cell; the minimum and maximum softening lengths in physical units are $369 \, \mathrm{pc}$ and $1.85 \, \mathrm{kpc}$, respectively. The cosmological parameters adopted are taken from the \citetalias{PlanckCollaboration2014}: $\mathrm{\Omega_m} = 0.307$, $\mathrm{\Omega_b} = 0.048$, $\mathrm{\Omega_\Lambda} = 0.693$, and a Hubble constant of $H_0 = 100 \, h \, \mathrm{km \, s^{-1} \, Mpc^{-1}}$, where $h = 0.6777$. The rest of the input parameters of the code are identical to those used in the Auriga Project \citep{Grand2017}.

As we show in the next section, our simulation, referred to as Au6\_MOL, produces a galaxy that, at $z = 0$, has a spiral structure, with an extended, rotationally supported disc and a bulge, in line with the expectations for the formation of a bulge-disc system in a galaxy, which has evolved quietly over several gigayears. This is consistent with Au6 being a galaxy with a quiet merger history and no significant merger events in its recent history (see \citealt{Grand2017}).

\section{Formation of molecular hydrogen and evolution of the gas and stellar phases}
\label{sec:results}

\begin{figure}[ht]
    \centering
    \includegraphics[width=8.8cm]{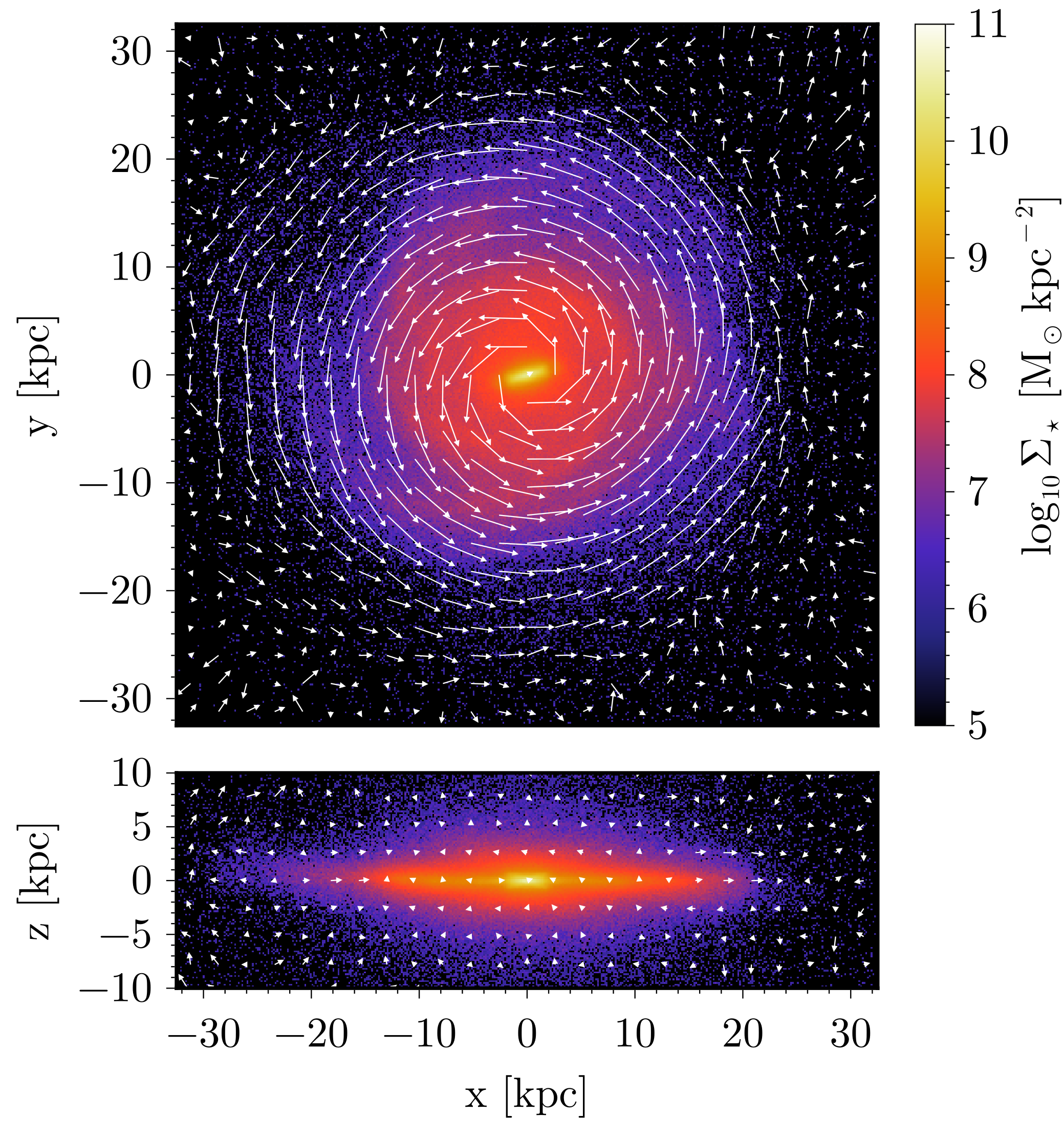}
    \caption{Projected stellar mass distributions, in face-on and edge-on views, for the simulation Au6\_MOL at $z = 0$. The white arrows indicate the corresponding velocity field.}
    \label{fig:stellar_map_Au6}
\end{figure}

\begin{figure}[ht]
    \centering
    \includegraphics[width=8.8cm]{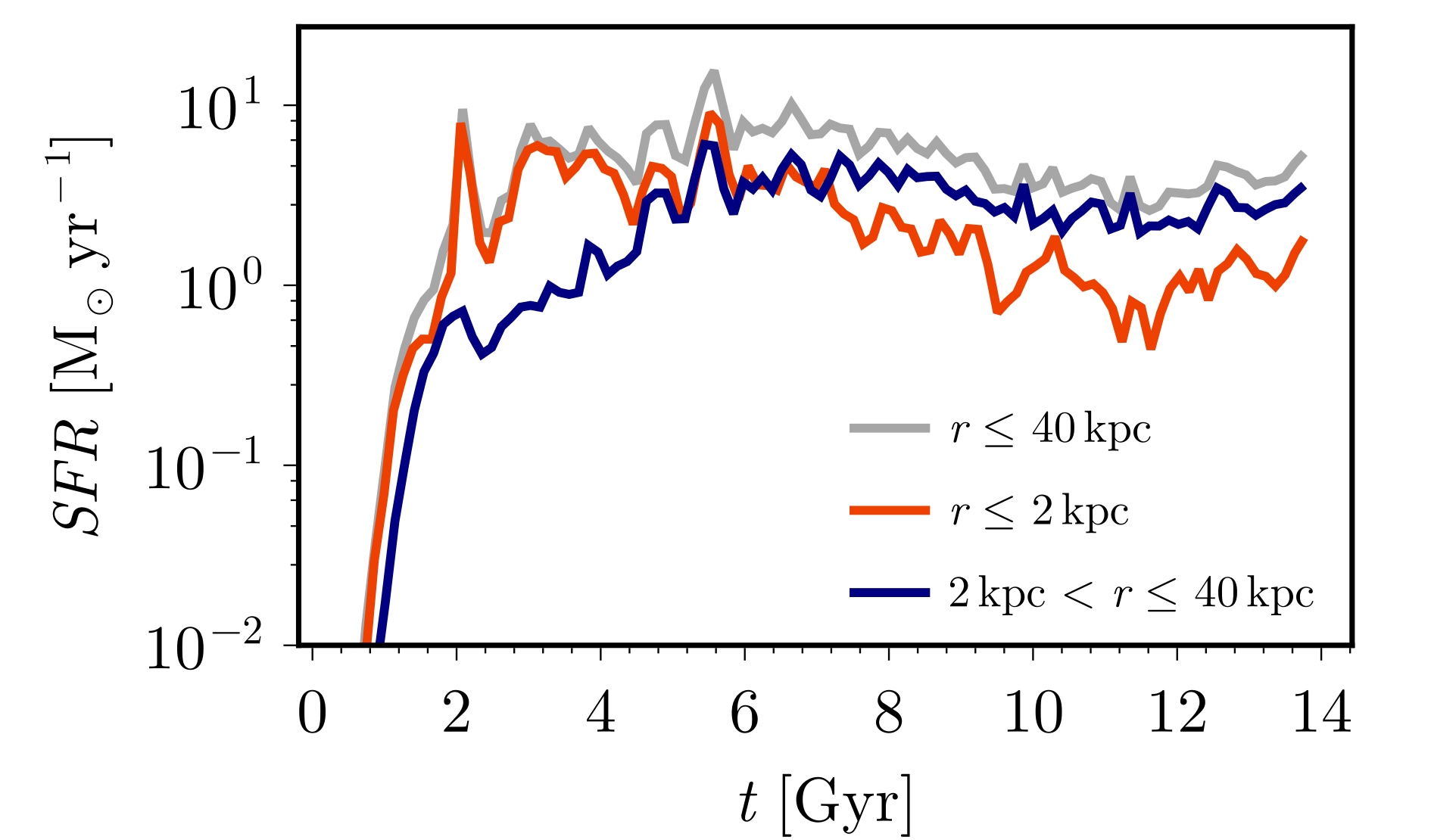}
    \caption{Evolution of the SFR for the simulation Au6\_MOL, computed at $z = 0$. The red and blue lines show the contribution of stars with $r \le 2 \, \mathrm{kpc}$ (mostly bulge stars) and $2 \, \mathrm{kpc} < r \le 40 \, \mathrm{kpc}$ (disc stars), respectively.}
    \label{fig:sfr_Au6}
\end{figure}

The final stellar distribution of the simulated galaxy can be observed in Fig.~\ref{fig:stellar_map_Au6}, where we show stellar density maps for the face-on and edge-on projections (the $z$-axis is aligned with the stellar angular momentum). Au6\_MOL presents a disc-like morphology, where ordered rotation (evident from the velocity field included in this figure) is observed up to a radius of about $25 \, \mathrm{kpc}$. The final morphology of the galaxy is the result of its SF history which, as shown in Fig.~\ref{fig:sfr_Au6}, grows rapidly during the first $\mathrm{Gyrs}$ of evolution, reaching a maximum at about $\sim \! 6-8 \, \mathrm{Gyr}$, and declining smoothly up to $z = 0$. The blue and red lines show separately the contributions of stars that, at $z = 0$, are in the bulge and disc regions, respectively (separated here only by the stellar radius). At $z = 0$, the SFR of the galaxy is $5.4 \, \mathrm{M_\odot \, yr^{-1}}$, and its final stellar mass is $6.5 \times 10^{10} \, \mathrm{M_\odot}$.

\begin{figure}[ht]
    \centering
    \includegraphics[width=8.8cm]{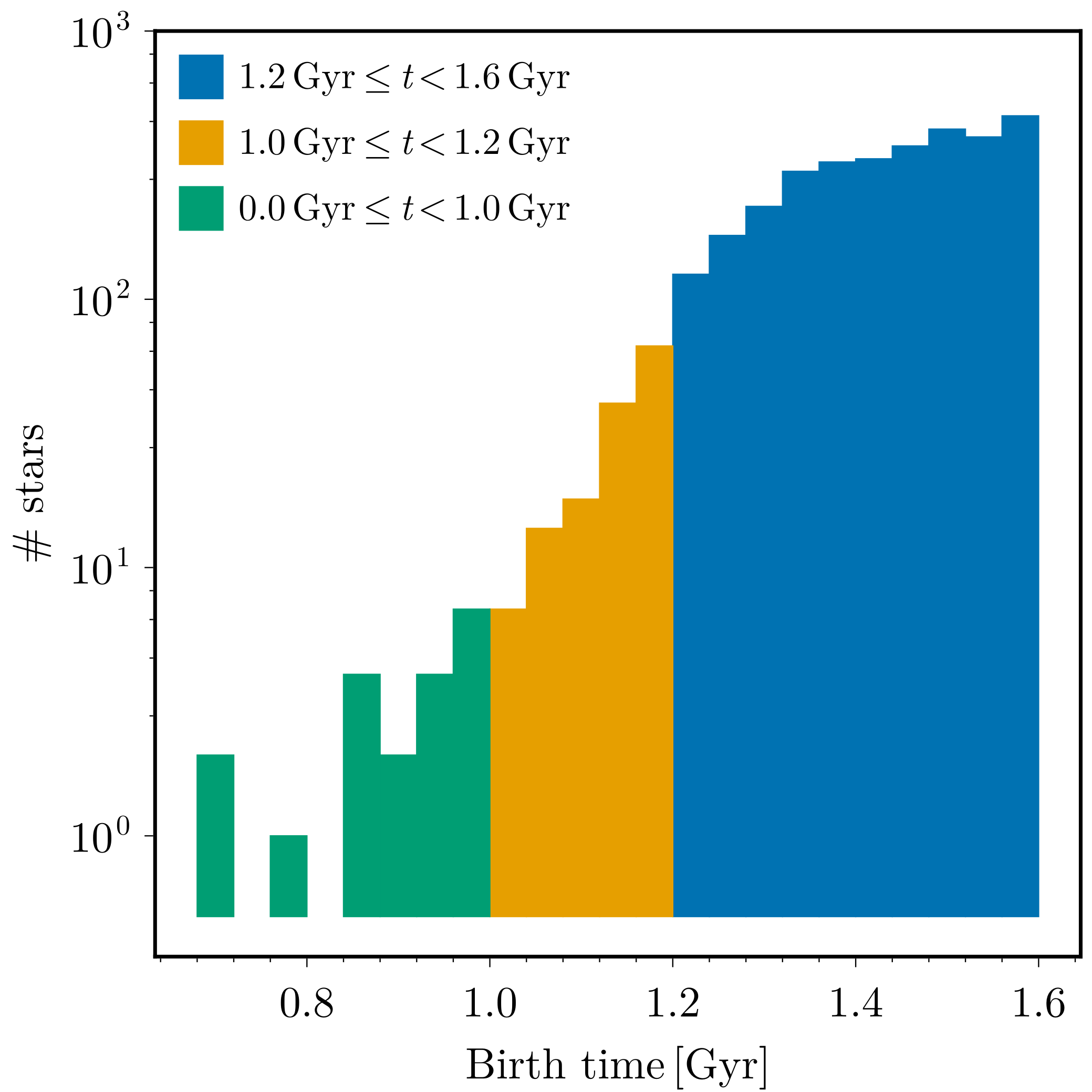}
    \includegraphics[width=8.8cm]{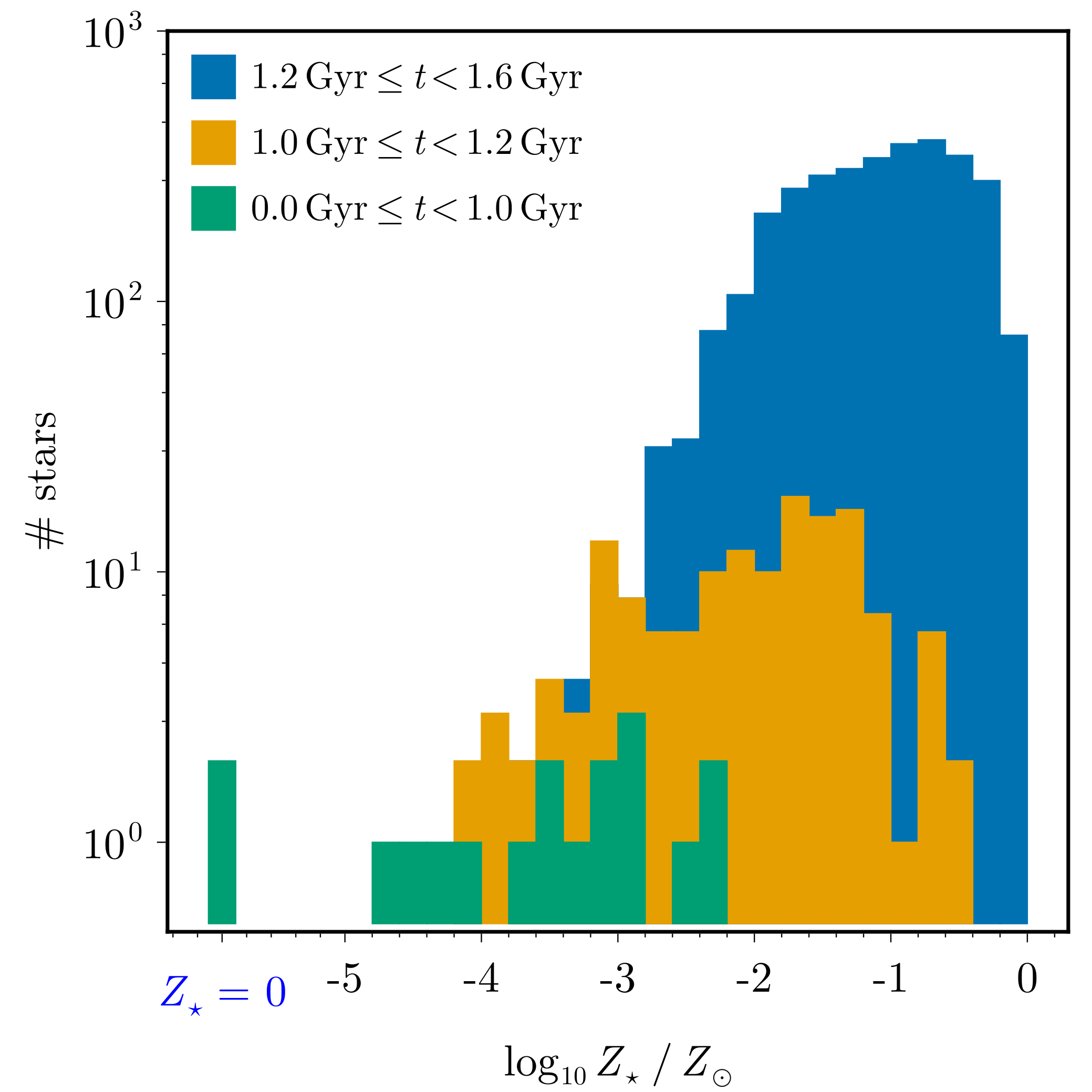}
    \caption{{\it Upper  panel}: Birth time distribution of the stars in the main sub-halo of simulation Au6\_MOL at $1.6 \, \mathrm{Gyr}$. The colours indicate different birth time ranges. {\it Lower panel}: Corresponding distribution of stellar metallicities. Metal-free stars are included artificially in the leftmost bin. Metal-free stars are only formed during the first time bin ($0.0 \, \mathrm{Gyr} \leq t < 1.0 \, \mathrm{Gyr}$).}
    \label{fig:bt_histogram_1Gyr}
\end{figure}

In our model, we assume that molecular gas is required to form stars, and thus the balance between $\mathrm{H}_2$ formation and destruction at early epochs is what sets the time at which stars start to form. Fig.~\ref{fig:sfr_Au6} shows that this occurs at $\sim \! 1 \, \mathrm{Gyr}$, indicating that the formation of molecular hydrogen has already started. The very first stars form from pristine gas, which (as we show in appendix~\ref{app:fixed_ics}) is highly inefficient at forming molecular hydrogen, even at high densities. However, after the collapse of the gas within the halo, high densities are achieved and a large amount of gas cells are eligible to form stars, entering our routine. Due to the stochastic nature of SF, some of these metal-free gas cells, even having a low molecular fraction, will be able to produce stars. The upper panel of Fig.~\ref{fig:bt_histogram_1Gyr} shows a histogram of the birth time of stars formed during the first $1.6 \, \mathrm{Gyr}$ of evolution. We can observe that only $20$ star particles form in the first $\mathrm{Gyr}$, indicating that molecular hydrogen has not yet been formed in a significant amount.

After the first $\mathrm{Gyr}$, a transition to significantly higher SF is found, with more than $3000$ star particles forming in the following $600 \, \mathrm{Myr}$. This transition is driven by the chemical enrichment of the ISM produced by the death of the first stars, which increases the production efficiency of star particles as molecular hydrogen forms more efficiently at higher metallicities (due to the inverse dependence of the condensation timescale on metallicity, see Eq. (\ref{eq:tau_cond})). The distribution of stellar metallicities for stars formed in the first $1.6 \, \mathrm{Gyr}$ of evolution is shown in the lower panel of Fig.~\ref{fig:bt_histogram_1Gyr} (the colours correspond to the same time intervals of the upper panel). From this plot we can observe that stars formed in the first $\mathrm{Gyr}$ are either metal-free ($10\%$) or with metallicities in the range $10^{-5} - 10^{-2.5} \, Z_\odot$. In the case of the other time bins, we find that none of the newly formed stars have $Z = 0$; and while some stars are still relatively metal-poor, even before $1.2 \, \mathrm{Gyr}$, many stars have already reached metallicities of at least $0.1 \, Z_\odot$, which is the typical metallicity where the transition to forming molecular hydrogen starts to be more efficient (see appendix~\ref{app:fixed_ics}).

\begin{figure}[ht]
    \centering
    \includegraphics[width=8.8cm]{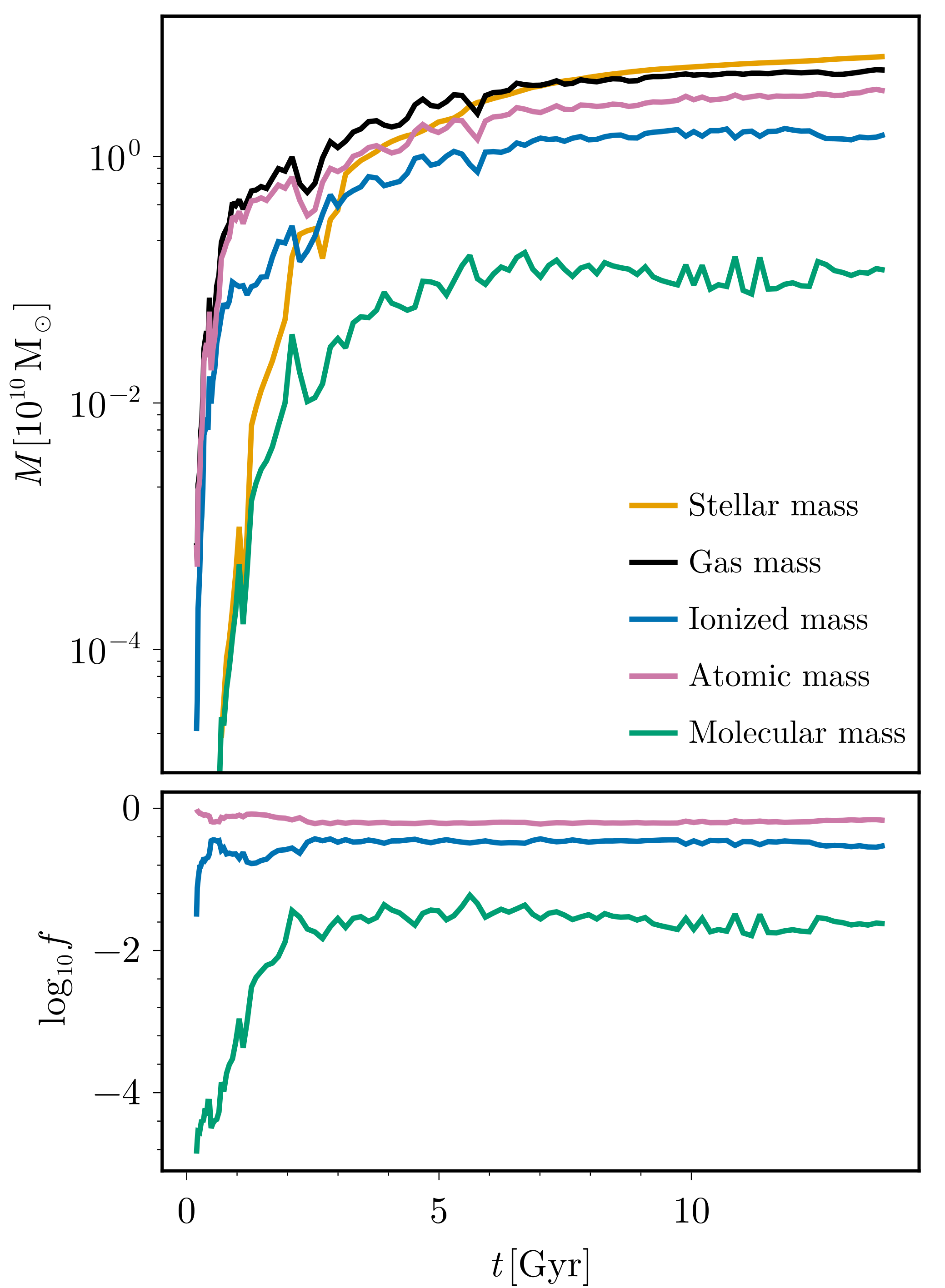}
    \caption{Evolution of the mass and relative fractions of the different gas phases for the main sub-halo of simulation Au6\_MOL up to a radius of $40 \, \mathrm{ckpc}$. We include the stellar mass and total gas mass for reference.}
    \label{fig:mass_and_fraction_evolution}
\end{figure}

As molecular hydrogen forms and SF is triggered, the simulated galaxy grows in a feedback-regulated regime, where the amount of gas in the different phases remains approximately constant, as we show in Fig.~\ref{fig:mass_and_fraction_evolution}\footnote{In this figure, we only consider material in the inner $40 \, \mathrm{ckpc}$ of the galaxy, in order to encompass the growth of the galaxy, and ignore gas in the halo.}. The fact that neither the molecular mass nor the mass in the atomic/ionised components vary significantly after the first $\sim \! 3 \, \mathrm{Gyr}$ of evolution, while SF continues, indicates an approximate equilibrium of mass exchange between the gas phases, consumption into stars, and mass return from SF. After the first $3 \, \mathrm{Gyr}$, the gas is distributed in the atomic, ionised, and molecular phases, respectively, with fractions of $\sim \! 63\%$, $\sim \! 33\%$, and $< 3\%$.

\begin{figure*}[ht]
    \centering
    \includegraphics[width=17cm]{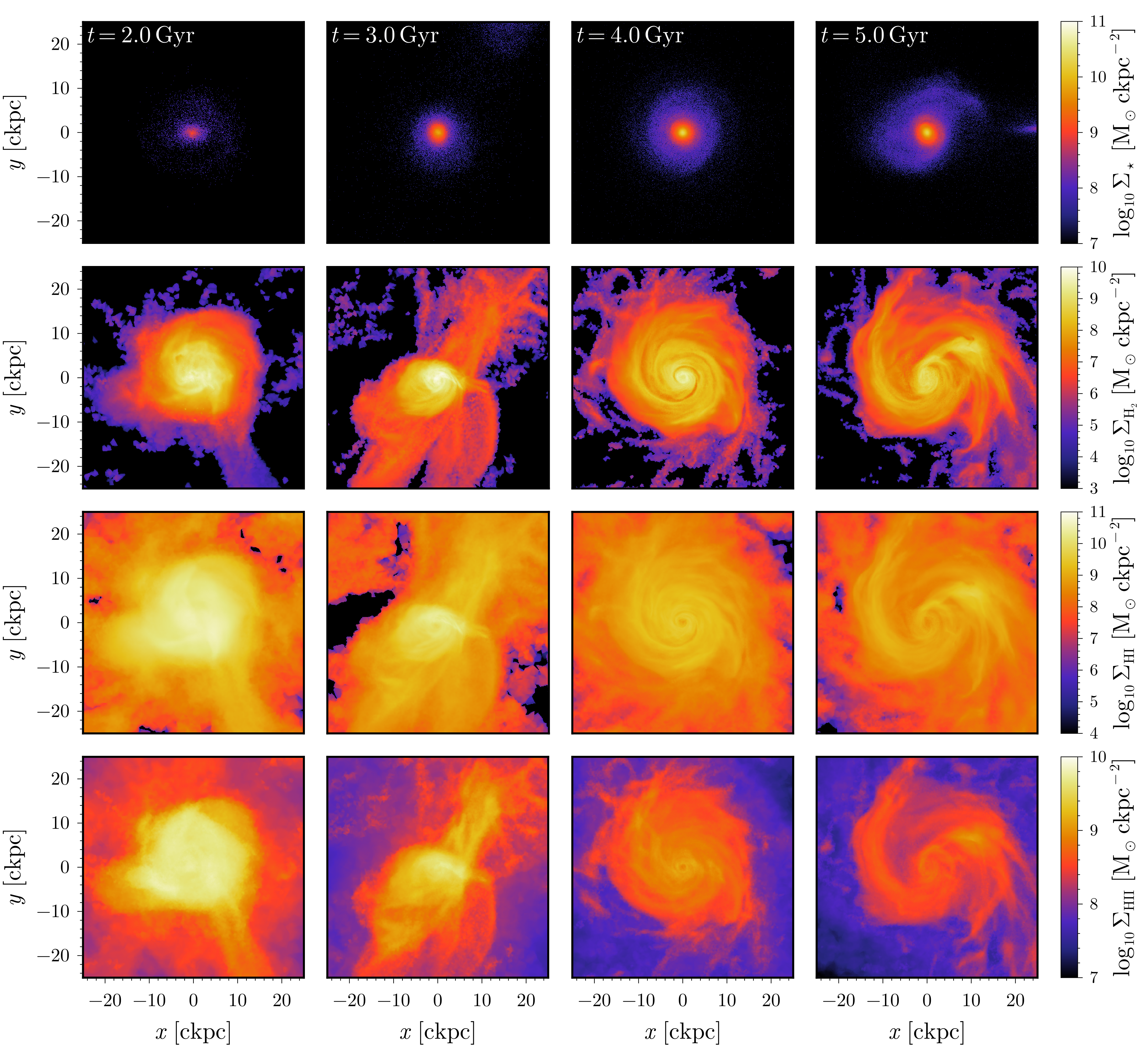}
    \caption{Projected mass density of the stars and the molecular, atomic, and ionised gas, at $2$, $3$, $4$, and $5 \, \mathrm{Gyr}$ for the simulation Au6\_MOL, in face-on views. The colour scales are fixed for each row to highlight the similarities and differences of the various distributions.}
    \label{fig:maps_vs_z}
\end{figure*}

The growth of the galaxy during the early phases of its formation, where our model is more relevant, is illustrated in Fig.~\ref{fig:maps_vs_z}, where we show the stellar and the molecular, atomic, and ionised gas distributions at $2$, $3$, $4$, and $5 \, \mathrm{Gyr}$ of evolution. While molecular gas is quite extended already at $2 \, \mathrm{Gyr}$, SF is only efficient in the central, most dense regions of the galaxy. A well-defined, extended disc structure in the gas is already present at $4 \, \mathrm{Gyr}$. This structure is maintained up to the present time, allowing the formation of the stellar disc. The three gas phases have a similar spatial structure at all times, in the innermost regions, although the atomic gas is more extended than the molecular and ionised phases. 

\begin{figure*}[ht]
    \centering
    \includegraphics[width=17cm]{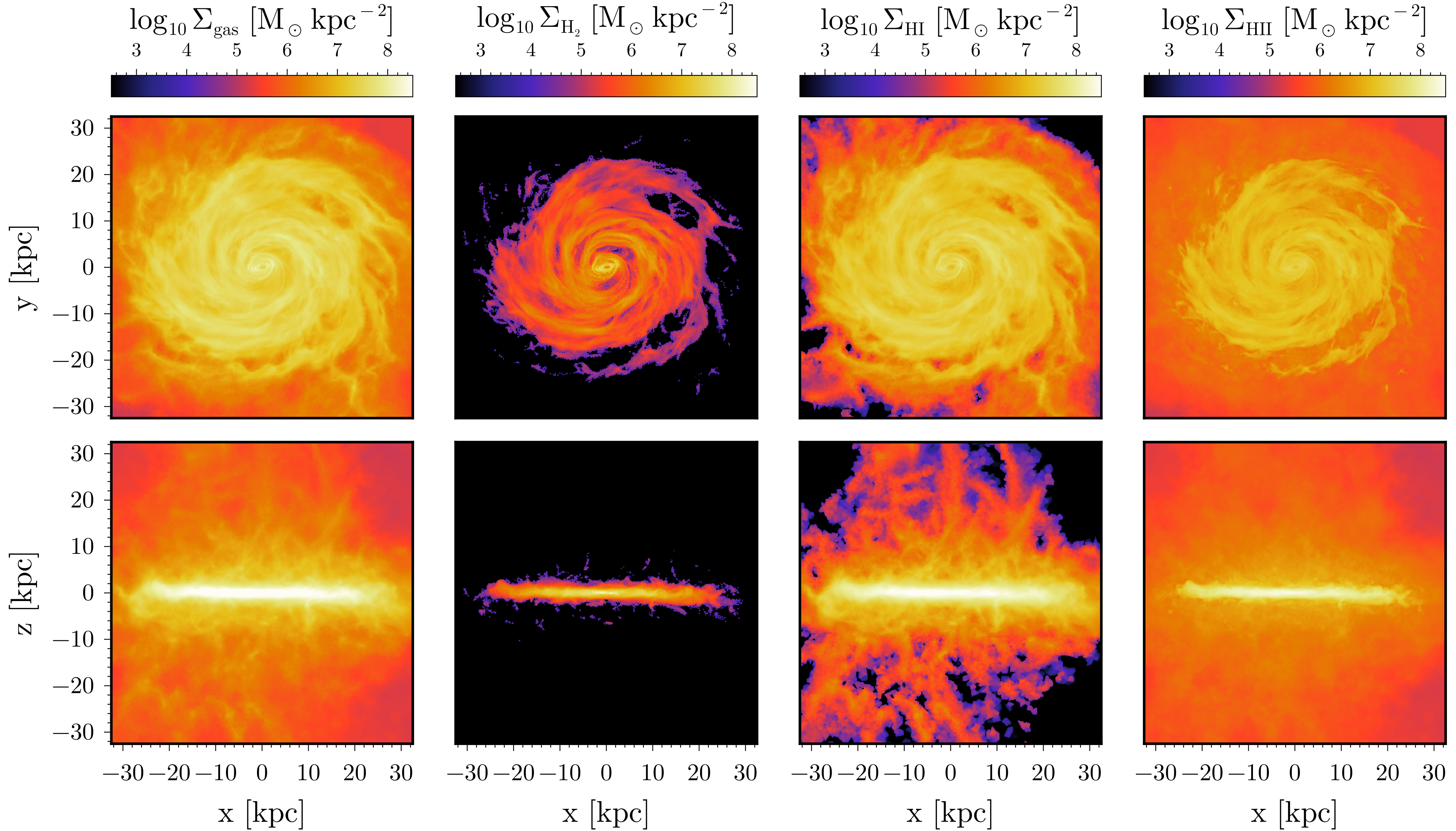}
    \caption{Projected mass density of the total, molecular, atomic, and ionised gas for the simulation Au6\_MOL at $z = 0$, in face-on (upper panels) and edge-on (lower panels) views. The colour scales are fixed to highlight the similarities and differences of the various distributions.}
    \label{fig:gas_maps_Au6}
\end{figure*}

After $5 \, \mathrm{Gyr}$, the galaxy grows in a smooth fashion with a slowly declining SFR and, by $z = 0$, the gas distributes itself in a disc-like configuration, as shown in Fig.~\ref{fig:gas_maps_Au6} (left-hand panels). The three gas phases (subsequent panels) are spatially correlated in the disc region (note that we kept the same colour scale in all cases, for a better comparison). Molecular gas is only present in the densest regions of the disc, while the atomic and ionised phases are more extended, both radially and vertically. The stellar mass and the mass gas in the different components, both in the inner $40 \, \mathrm{kpc}$ and up to the virial radius, are given in Table~\ref{tab:disc_sizes}. In the table we also include the corresponding values for the $R_{95}$ parameter, defined as the radius, which encloses $95\%$ of the total mass of any given component (see \citealt{Iza2022})\footnote{In the case of the gas, it is necessary to adopt a maximum radius to compute the $R_{95}$ of the disc, which we take as $40 \, \mathrm{kpc}$. This is because the gas extends up to the virial radius, and a non-negligible fraction is located outside the disc (this is not formally needed for the stellar component, as there are almost no stars outside the central region of the system. We checked that varying the maximum stellar radius in the range $40-60 \, \mathrm{kpc}$ changes the stellar $R_{95}$ by less than $1\%$.}. As evident when we compare Figs.~\ref{fig:gas_maps_Au6} and ~\ref{fig:stellar_map_Au6}, the gas disc has approximately twice the size of the stellar disc, which is confirmed by the values for $R_{95}$: $15.6 \, \mathrm{kpc}$ for the stars and $\sim \! 35 \, \mathrm{kpc}$ for the ionised gas phase. The $R_{95}$ value of the molecular gas ($\sim \! 20 \, \mathrm{kpc}$) is in line with the stellar size. 

\begin{table}[ht]
    \caption{Characteristic radius and mass in the different components.}
    \centering
    \def\arraystretch{1.5}
    \begin{tabular}{c|ccc|}
    \cline{2-4}
        & \multicolumn{3}{c|}{\textbf{Au6\_MOL}} \\ \cline{2-4} 
        & \multicolumn{1}{c|}{\begin{tabular}[c]{@{}c@{}}$R_{95}$\\ $\mathrm{[kpc]}$\end{tabular}} & \multicolumn{1}{c|}{\begin{tabular}[c]{@{}c@{}}$M_\mathrm{tot}$\\ $\mathrm{[10^{10} \, M_\odot]}$\end{tabular}} & \begin{tabular}[c]{@{}c@{}}$M_\mathrm{<40 \, kpc}$\\ $\mathrm{[10^{10} \, M_\odot]}$\end{tabular} \\ \hline
        \multicolumn{1}{|c|}{Stars} & 15.6 & 6.5 & 6.48 \\
        \multicolumn{1}{|c|}{Gas} & 32.0 & 7.83 & 5.03 \\
        \multicolumn{1}{|c|}{\HII} & 35.3 & 4.25 & 1.51 \\
        \multicolumn{1}{|c|}{\HI} & 30.3 & 3.46 & 3.4 \\
        \multicolumn{1}{|c|}{$\mathrm{H_2}$} & 20.1 & 0.12 & 0.12 \\ \cline{1-4}
    \end{tabular}
    \tablefoot{At $z = 0$, for simulation Au6\_MOL: $R_{95}$ is the radius that contains $95\%$ of the mass within a sphere of $r = 40 \, \mathrm{kpc}$, $M_\mathrm{tot}$ is the total mass of the main sub-halo, and $M_\mathrm{< \, 40 \, \mathrm{kpc}}$ is the mass within the inner $40 \, \mathrm{kpc}$. We show results for the stellar component, as well as for the total gas, and the ionised, atomic, and molecular phases.}
    \label{tab:disc_sizes}
\end{table}

\begin{figure}[ht]
    \centering
    \includegraphics[width=8.8cm]{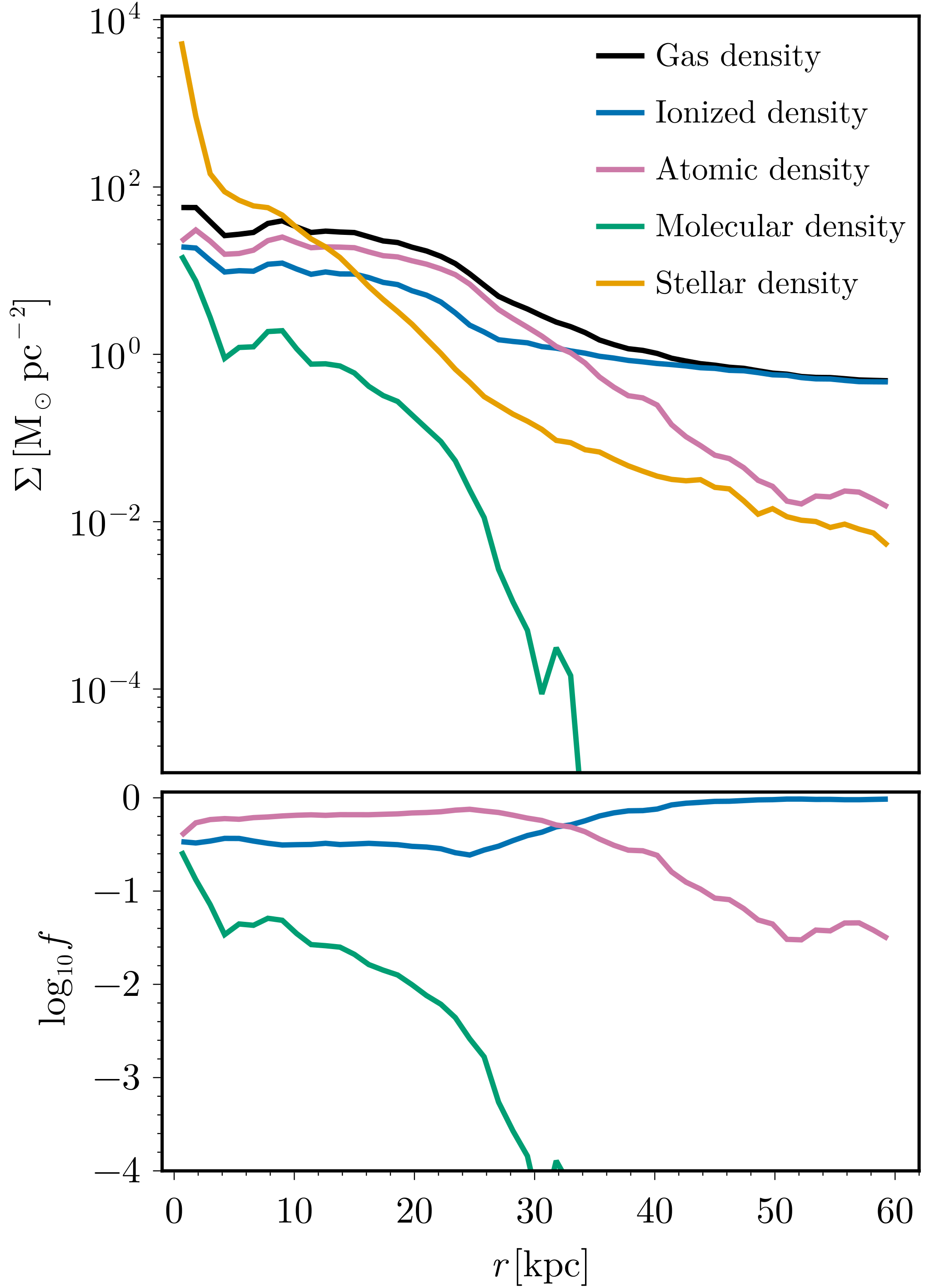}
    \caption{{\it Upper panel}: Projected mass density profiles for the gas and stellar components of the simulation Au6\_MOL at $z = 0$. We show separately the contributions of the ionised, atomic, and molecular phases, and include the stellar component, for comparison. {\it Lower panel}: Radial profiles of the global mass fraction of the ionised, atomic, and molecular gas components, normalised to the total mass of gas.}
    \label{fig:density_profile_gas_Au6}
\end{figure}

The way in which the gas is distributed spatially in the different phases can be better seen from the surface density profiles shown in Fig.~\ref{fig:density_profile_gas_Au6} and the profiles of the relative fractions of the three gas phases. From this plot it is clear that the atomic phase largely dominates the gas mass in the disc region and that the transition to the halo, where ionised gas starts dominating, occurs at $\sim \! 32 \, \mathrm{kpc}$. The molecular gas is always a small fraction of the gas, although it is worth noting that this component is constantly being turned into stars. As shown in Table~\ref{tab:disc_sizes} the relative proportion between atomic and ionised gas in the disc region is approximately 2.3:1, and the molecular gas comprises less than $3\%$ of the total gas mass.

\section{Predictions for the SF law and SF efficiency}
\label{sec:ks_law_epsilon_ff}

\begin{figure}[ht]
    \centering
    \includegraphics[width=8.8cm]{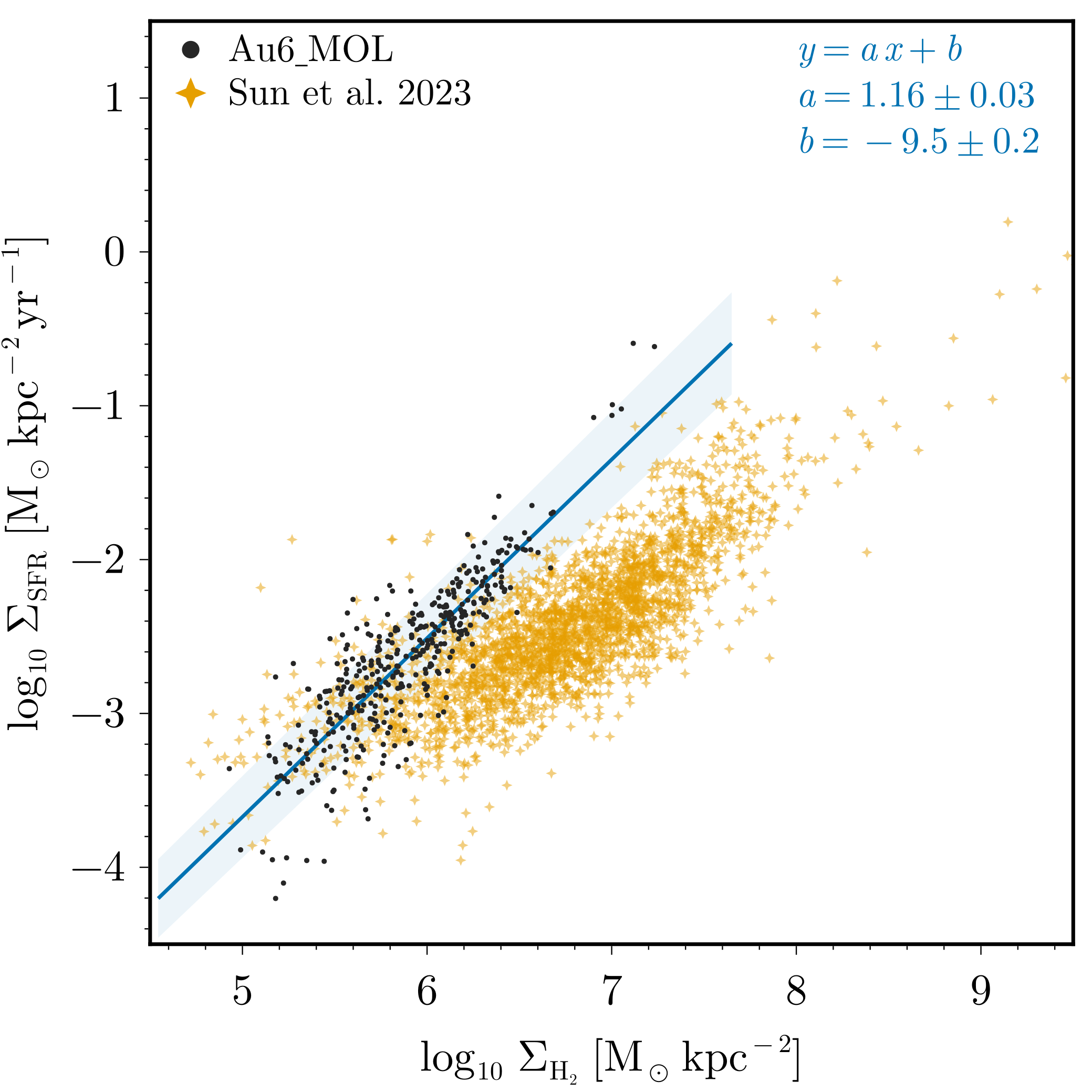}
    \caption{Molecular Kennicutt–Schmidt relation, at $z = 0$, for our simulations Au6\_MOL (black dots). Each point corresponds to a $1.5 \, \mathrm{kpc} \times 1.5 \, \mathrm{kpc}$ parcel of a $30 \, \mathrm{kpc}$ square region centred on the main sub-halo. We also include a linear fit (blue line) for our data points and the observational values of \cite{Sun2023} (orange).}
    \label{fig:mKS}
\end{figure}

To test our SF model, in this Section we investigate its predictions for the molecular Kennicutt–Schmidt (mKS) relation, and for the SFEs. As shown in Fig.~\ref{fig:mKS}, the predicted mKS relation for our simulation Au6\_MOL at $z = 0$ is obtained using young stellar particles (up to stellar ages of $200 \, \mathrm{Myr}$). Even though in our model the KS law is not imposed (see Section~\ref{sec:simulations}), we find a well-defined correlation between the $\mathrm{H}_2$ and SFR surface densities. Our data points cover more than 2 orders of magnitude in both axes, with the largest $\mathrm{H}_2$ densities corresponding to the inner regions of the galaxy. The best-fit parameters associated with a linear fit to the data are also included in the figure. 

In the figure we also include the resolved observations of \citet{Sun2023}, who analysed the mKS relation across 80 nearby star-forming spiral galaxies at a resolution of $1.5\, \mathrm{kpc}$. In both our model and the observations, the SFR surface density, $\Sigma_{\mathrm{SFR}}$, increases with molecular gas surface density, $\Sigma_{\mathrm{H}_2}$, following a power-law trend. While the general trend obtained in our simulation is similar to the observational findings, we find a higher slope, of $1.16 \pm 0.03$, which lies within the upper end of the observational range reported by \citet{Sun2023} ($0.88$ to $1.21$, depending on methodology). In particular, the simulation predicts higher values of $\Sigma_{\mathrm{SFR}}$ at $\Sigma_{\mathrm{H}_2} \gtrsim 6\, \mathrm{M_\odot\, pc^{-2}}$. It is possible that our simplified treatment of gas shielding is partly responsible for this discrepancy; we will explore the effects of self-shielding on the mKS relation in a forthcoming paper.

The predicted distribution of SF efficiencies (SFE) for gas cells can also be used as a test for the model, to quantify the degree of agreement with available measurements. We adopt the usual parametrisation of the SFR of gas elements, i.e.
\begin{equation}
    \dot{\rho}_\mathrm{\star} = \epsilon_\mathrm{ff} \, \frac{\rho_\mathrm{g}}{t_\mathrm{ff}} \, , 
\end{equation}
where $\rho_\mathrm{g}$ is the gas density and $t_\mathrm{ff}$ is the free-fall time of the cloud, to obtain $\epsilon_\mathrm{ff}$, a quantity that can be physically interpreted as the cloud SFE per free-fall time. In the simulations, we can compute $\epsilon_\mathrm{ff}$ for each cell, which represents the average SFE per free-fall time at the resolved scale. In fact, we cannot only obtain the predicted $\epsilon_\mathrm{ff}$ distributions at all times, but we can also study possible dependencies on properties such as the gas density and metallicity, which are the two most important parameters of our SF prescription that correlate with the SFR (see appendix~\ref{app:fixed_ics}). 

\begin{figure*}[ht]
    \centering
    \includegraphics[width=17cm]{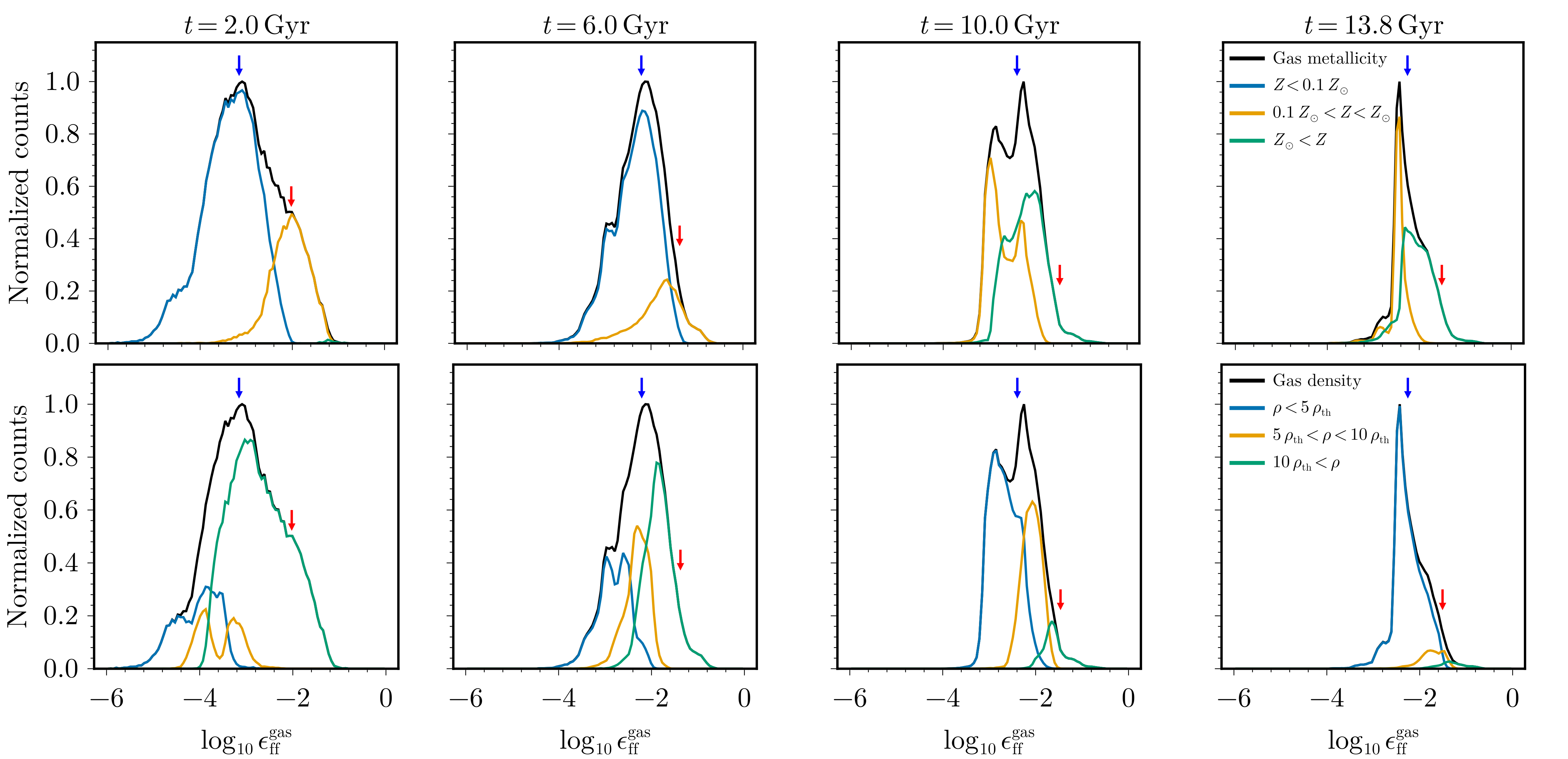}
    \caption{Distribution of the SFE per free fall time of the gas in the disc ($r < 40 \, \mathrm{ckpc}$) of simulation Au6\_MOL, at $2, 6, 10$, and $13.8 \, \mathrm{Gyr}$. The first row shows the distribution separated by metallicity ranges (the colour lines show the ranges while the black lines show the full distribution), and the second row shows the distribution separated by density ranges ($\rho_\mathrm{th}$ in the legend is the density threshold for SF). The blue arrows indicate the median of the total distribution (black line), while the red arrows indicate the median of the distribution for gas that has been transformed into stars (this distribution is not shown).}
    \label{fig:epsilon_ff_distributions}
\end{figure*}

The distribution of $\epsilon_\mathrm{ff}$ in our simulation Au6\_MOL, at different times, is shown in Fig.~\ref{fig:epsilon_ff_distributions}. As discussed in the previous section, our SF model has a higher impact at early times, in the low-metallicity regime of the gas. The upper and lower rows in the figure include, respectively, the results obtained for three different ranges in metallicity and density. Several important results are revealed by this figure. First, the values of the SFE cover a wide range between $\sim \! 0.001\%$ to $\sim \! 10\%$. As expected, the lowest values of $\epsilon_\mathrm{ff}$ and the largest spread is found at $t = 2 \, \mathrm{Gyr}$. Second, there is no variation in the maximum possible values of $\epsilon_\mathrm{ff}$, which are always of the order of $10\%$. Third, the median values of the distributions (indicated in the figure with the blue arrows) are $0.4-0.6\%$ for all times except for $t = 2 \, \mathrm{Gyr}$. Finally, the SFE increases with increasing metallicity and density (coloured lines). As discussed previously, in our model the SFR per cell depends on various parameters in a complex way, including the density and the metallicity (see also next section). 

We note that the increasing trend of SFE with metallicity seen in in Fig.~\ref{fig:epsilon_ff_distributions} is not a prediction that emerges independently from the model's dynamics, but rather a natural consequence of our assumptions regarding $\mathrm{H}_2$ formation. Specifically, the model assumes that molecular hydrogen formation is more efficient in higher-metallicity environments due to more abundant catalytic sites, leading to higher molecular fractions and thus higher SFE. While the absolute values of SFE are a prediction of the model, the qualitative dependence on metallicity reflects this built-in assumption.

We can also calculate the values of $\epsilon_\mathrm{ff}$ for gas cells that produced star particles, right before this occurs. In this case, we find that, as expected, they belong to the high tail of the distributions as they have higher probability of forming stars. The median values of the distributions for gas cells that produced stars are indicated by the red arrows in Fig.~\ref{fig:epsilon_ff_distributions}, and are between $3-4\%$ at all times, except for $t = 2 \, \mathrm{Gyr}$ where the median value is $0.9\%$. 

These results for the SFE per free-fall time in our simulation agree very well with observations from giant molecular clouds, which suggest that $\epsilon_\mathrm{ff}$ varies from $0.1\%$ to $10\%$ (e.g. \citealt{Krumholz2019,Schinnerer2024}), and resolved measurements in nearby star-forming spiral galaxies and in the MW, with values of the order of $\sim \! 0.7\%$ \citep{Utomo2018,Sun2023}. Moreover, both the variations and values obtained with our model are consistent with those found in SF simulations in a turbulent ISM \citep{Semenov2016}, when averaged over the resolved scales.

\section{Discussion}
\label{sec:discussion}

Due to the complex cycle between gas, SF, and feedback in galaxies, the development of a physically motivated model for SF that works at the resolved scales (ideally capturing also our knowledge on smaller, unresolved scales) is not an easy task. Moreover, the tuning of the input parameters associated with SF cannot be done separately from other parameters of the model, most notably those related to feedback, which provides SF regulation and sets the SF levels. Our model has input parameters within the system of equations, but most of them can be obtained from observations or theoretical considerations (see Section~\ref{subsec:parameters}). As a first test to our SF model, we compared the molecular KS law and the distribution of SFE per free-fall time with several observations.

In this section, we compare the results of our model to those of two other runs that use alternative SF laws, commonly adopted in zoom-in simulations of galaxy formation. The differences between the three models are in the estimation of the SFR of gas cells, which alters the probability of forming stars, i.e.,
\begin{equation}
    p = 1 - \exp\left(- F_s\right) \, ,
\end{equation}
where $F_s$ represents the fraction of stellar mass in the gas cell that was formed during time step $\Delta t$ (see Section~\ref{subsec:model_equations}). The assumptions for $F_s$ in the different models are as follows:
\begin{itemize}
    \item Au6\_MOL assumes that $F_s = f_s$ where $f_s$ is the stellar mass fraction obtained using our system of equations.
    \item Au6\_STD uses the standard \texttt{AREPO} formulation (as adopted in the Auriga Project) where $F_s$ depends on the cold mass fraction ($x$), and the assumed SF timescale ($t_*$, with a maximum SF timescale of $t^*_0 = 2.2 \, \mathrm{Gyr}$), as described in Section~\ref{subsec:model_equations}:
    \begin{equation} 
        F_s = \frac{x}{t_*} \, \Delta t \, .
    \end{equation}
    \item Au6\_BLT uses a molecular-based SF probability, which depends on an estimated molecular fraction $f_m$ per cell,  obtained using the observationally motivated relation of \cite{Blitz2006}, i.e.:
    \begin{equation}
        f_m = \frac{1}{1 + P_0 / P} \, ,
    \end{equation}
    where $P$ is the gas pressure of the cell and $P_0 / k = 3.5 \times 10^4  \, \mathrm{K} \, \mathrm{cm^{-3}}$.
    The $F_s$ value in this model is 
    \begin{equation}
        \label{eq:blitz}
        F_s = f_* \, \frac{f_m}{t_\mathrm{ff}} \, \Delta t \, ,
    \end{equation}
     where $f_*$ is an efficiency parameter and $t_\mathrm{ff}$ is the free-fall time of the cell. Following \cite{Murante2010,Murante2014}, who adopt a similar approach, we take $f_* = 0.02$ (see also \citealt{Valentini2022}).
\end{itemize}

\begin{figure}[ht]
    \centering
    \includegraphics[width=8.8cm]{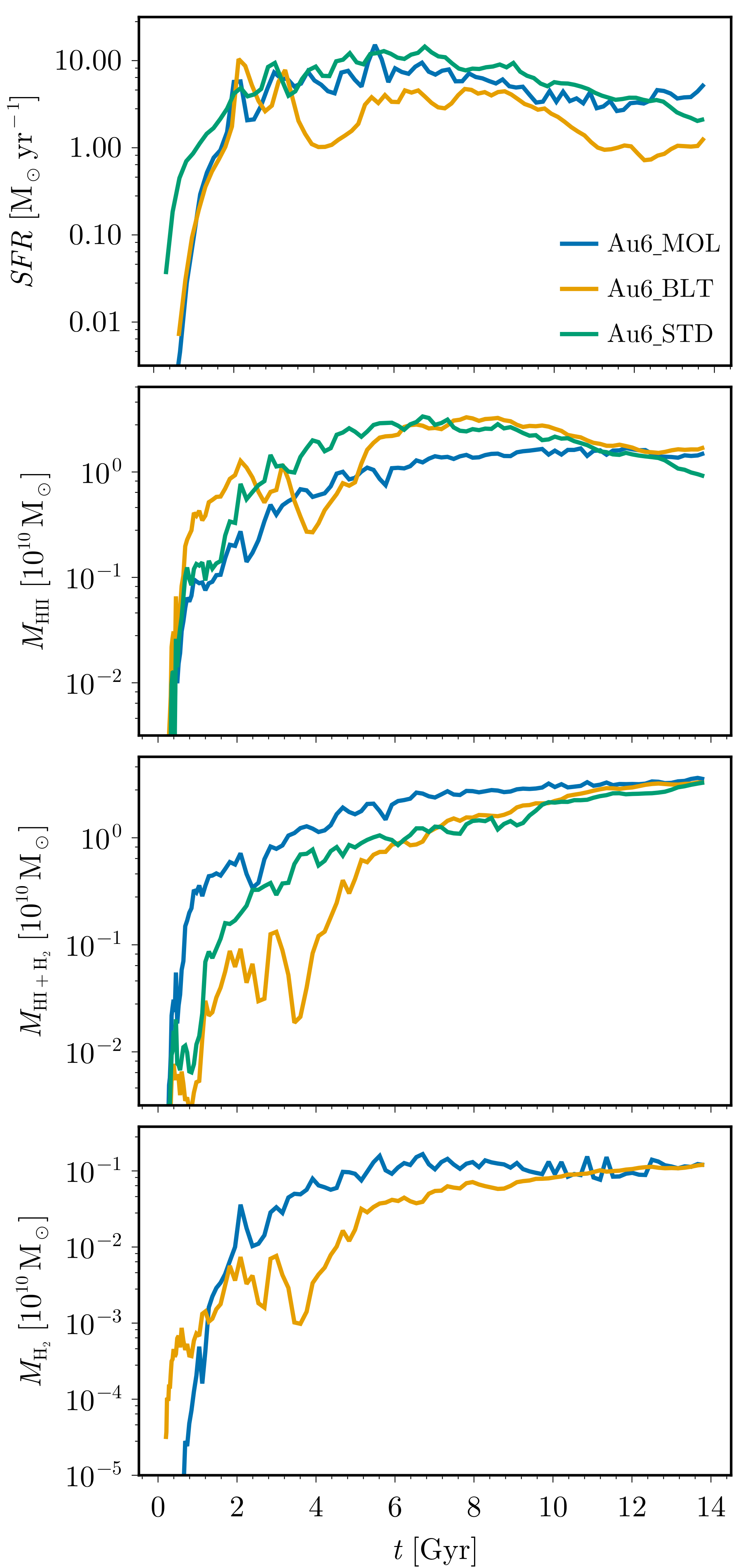}
    \caption{Comparison of the SFR and evolution of the gas mass in the ionised, neutral, and molecular components between the three simulations with different SF laws: Au6\_MOL (our model), Au6\_BL (an alternative $\mathrm{H}_2$-based SF law), and Au6\_STD (the standard \texttt{AREPO} implementation, where SF is linked to the total gas density). Only gas cells within a sphere with $r = 40 \, \mathrm{ckpc}$ are considered.}
    \label{fig:sfr_and_mass_evolution_comparison}
\end{figure}

\begin{figure*}[ht]
    \centering
    \includegraphics[width=17cm]{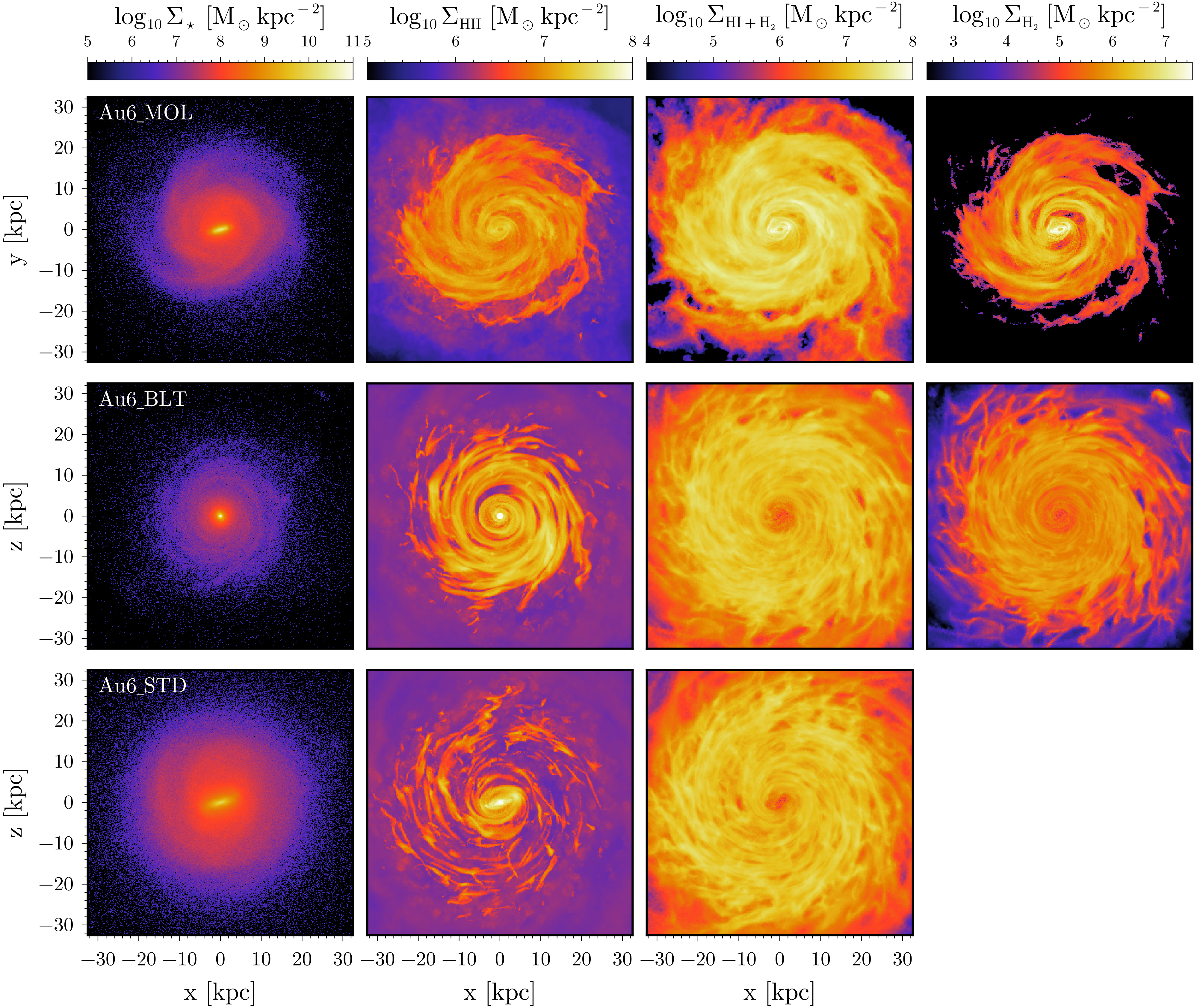}
    \caption{Projected mass densities of the stellar, ionised, neutral, and molecular gas components at $z = 0$ for the three simulations with different SF laws: Au6\_MOL (our model), Au6\_BL (an alternative $\mathrm{H}_2$-based SF law), and Au6\_STD (the standard \texttt{AREPO} implementation, where SF is linked to the total gas density).}
    \label{fig:density_distribution_comparison}
\end{figure*}

The comparison between the three simulations are summarised in Fig.~\ref{fig:sfr_and_mass_evolution_comparison} where we show the evolution of the SFR and of the gas mass in the different components and in Fig.~\ref{fig:density_distribution_comparison} where we show the $z = 0$ distributions of the stars and the ionised, atomic, and molecular gas (note that there is no estimation of the molecular mass for Au6\_STD). Looking at the top panel of Fig.~\ref{fig:sfr_and_mass_evolution_comparison}, the SFRs of the Au6\_MOL and Au6\_STD models are very similar, while Au6\_BLT lies below. More important is the fact that the two models using $\mathrm{H}_2$-dependent SF laws show a delay in the onset of SF of about $0.5 \, \mathrm{Gyr}$ with respect to Au6\_STD, which results from the need to form molecular gas first before forming stars. In our model, this is a consequence of the low enrichment levels of the ISM at these early times. The evolution of the mass in the ionised, neutral (i.e. atomic plus molecular in the Au6\_MOL and Au6\_BLT cases), and molecular phases also presents some differences between the models, even though the $z = 0$ values are very similar in all cases. We also observe that Au6\_BLT produces a lower amount of molecular gas in comparison with our model, which explains the observed differences in the SFRs. 

As it can be seen in Fig.~\ref{fig:density_distribution_comparison}, the galaxies end up with similar morphologies but somewhat different disc sizes for the stellar and gas components. Interestingly the molecular gas is more extended in Au6\_BLT than in Au6\_MOL, whereas the corresponding stellar discs show the opposite trend. In contrast to our model, where the amount of $\mathrm{H}_2$ depends on the density and metallicity, in Au6\_BLT the $\mathrm{H}_2$ fraction depends only on the gas pressure. The structure of the gas in the different components also shows some differences that are more evident in the ionised phase (which is consistent with the results of \citealt{Valentini2022}, but see also \citealt{Christensen2012}).

\begin{figure*}[ht]
    \centering
    \includegraphics[width=17cm]{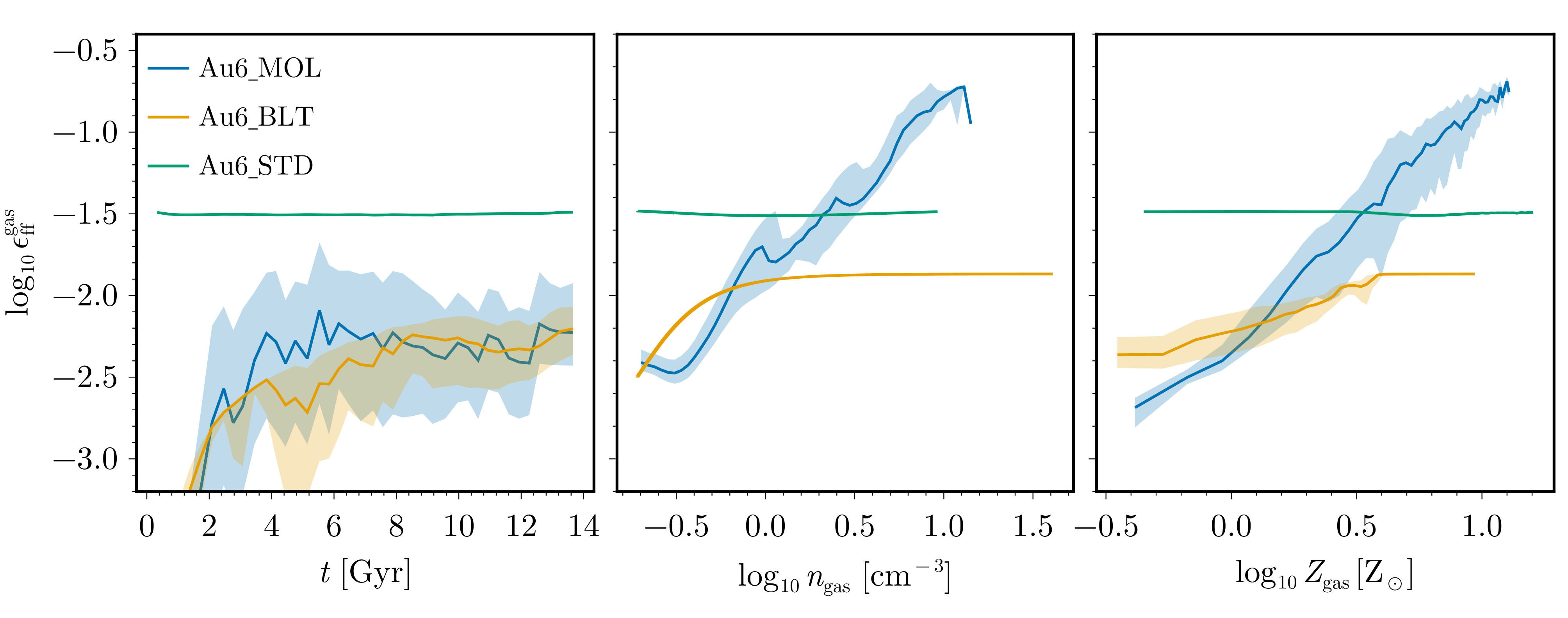}
    \caption{ {\it Left:} Time evolution of the median of $\epsilon_\mathrm{ff}$ for the three simulations, together with the median absolute deviation (shades). {\it Middle:} Correlation between the SFE parameter and the gas density at $z = 0$. {\it Right:} Correlation between the SFE parameter and the gas metallicity at $z = 0$.}
    \label{fig:epsilon_ff_comparison}
\end{figure*}

In the left-hand panel of Fig.~\ref{fig:epsilon_ff_comparison} we compare the evolution of the median SFE per free-fall time ($\epsilon_\mathrm{ff}$) for gas cells, also including the shaded range from the $25\%$ to the $75\%$ percentiles of the data. In the case of Au6\_STD, $\epsilon_\mathrm{ff}$ depends on the cold mass fraction and the assumed SF timescale (Section~\ref{subsec:model_equations}), while in Au6\_BLT it depends on the fraction of molecular hydrogen and the assumed SF efficiency parameter ($f_*$ in Eq. (\ref{eq:blitz})). In Au6\_STD, $\epsilon_\mathrm{ff}$ stays approximately constant during the whole evolution, at a value of $3\%$. This results from the fact that there are no significant variations in the cold gas fraction of star-forming gas cells (note that, although unnoticeable at the scale of the plot, there is a non-zero spread in $\epsilon_\mathrm{ff}$). In contrast to Au6\_STD, the two models where the SF law is based on $\mathrm{H}_2$ show a noticeably evolution of the typical SFE, as well as a large spread, most notably in the case of our Au6\_MOL model. 

In the middle- and right-hand panels of Fig.~\ref{fig:epsilon_ff_comparison} we further explore how the SFEs depend on the properties of the gas, showing $\epsilon_\mathrm{ff}$ as a function of gas density and metallicity, at $z = 0$ (similar results are found for other redshifts). Both Au6\_MOL and Au6\_BLT present a positive correlation between the SFE and density/metallicity, and these correlations are significantly stronger in Au6\_MOL compared to Au6\_BLT, even though both simulations show similar SFEs, particularly in the last $6 \, \mathrm{Gyr}$ of evolution (left-hand panel of this figure). This is explained by the fact that in Au6\_BLT, there is a direct relation between $\epsilon_\mathrm{ff}$ and gas density, because in this model the molecular fraction only depends on the pressure; and the dependence on metallicity results from correlations between the gas density and the metallicity. 

We note that, additionally to the use of the molecular KS law, the distribution, evolution, and correlations of the SFE per free-fall time predicted by different models could be useful as a test for SF routines in simulations where the structure of the ISM is not resolved. This might be particularly useful in the case of our model, where the SF routine can be easily adapted to link SF to the molecular gas, the atomic gas (likely precursor of SF at high redshift), or a combination of both. 

\section{Conclusions}
\label{sec:conclusions}

In this work we presented a model to describe the various components of hydrogen -- ionised, atomic, and molecular -- in the context of numerical simulations of galaxy formation and analysed the effects of linking SF to the molecular phase. Our implementation aims at providing a more realistic model for SF in simulations, particularly to allow a varying SFE of the gas cells, which can be linked to the local properties of the gas. The model is based on the formulation of \cite{Millan-Irigoyen2020} and grafted onto the cosmological, magnetohydrodynamical moving-mesh code \texttt{AREPO}.
 
In our sub-grid model, we assumed that each gas cell is composed of ionised, atomic, and molecular gas, as well as stars. These components exchange mass through recombination (ionised to atomic), condensation (atomic to molecular), SF (molecular to stellar), photodissociation (molecular to atomic), photoionisation (atomic to ionised), and mass return (stellar to ionised). While the modelling is relatively simple, it allows us to account for variations in the corresponding efficiencies and typical timescales of the various processes, according to the properties of the local ISM. In our model, the SFR per gas cell is assumed to be directly proportional to the amount of $\mathrm{H}_2$, with a typical timescale that depends on gas density, and with no additional factors (i.e. we do not introduce any particular SFR efficiency parameter). Although we change the SF prescription to link the SFR to the amount of $\mathrm{H}_2$, we adopt the standard \texttt{AREPO} modelling for primordial and metal-line cooling, reionisation from a uniform ultraviolet background field, magnetic fields, energy feedback from Type II supernovae, and chemical enrichment via supernovae II and Ia and AGB stars. However, in this work, we switched off the effects of active galactic nuclei, as their effects are subdominant for MW-mass galaxies.

We applied our model to a cosmological zoom-in IC of the formation of a MW-mass galaxy, taken from the Auriga Project \citep{Grand2017}. In order to compare our model with other prescriptions, we also ran two additional simulations, using the standard SF law of \texttt{AREPO}, and an $\mathrm{H}_2$-based SF model where the abundance of molecular hydrogen is obtained from an observationally motivated relation with the local gas pressure \citep{Blitz2006}.

Our model predicts the formation of a spiral-like galaxy, with a young, rotationally supported disc component, and a small bulge. The gas disc is twice as large as the stellar one and composed mainly of atomic ($\sim \! 63\%$) and ionised ($\sim \! 33\%$) material. The molecular gas is always a small fraction of the total gas ($\lesssim \! 3\%$), which results from a rapid consumption of $\mathrm{H}_2$ into stars. The relative fractions of atomic and ionised material in the disc remain approximately constant during the whole evolution, even at times between $1 \, \mathrm{Gyr}$ and $3 \, \mathrm{Gyr}$ where the SFR changes rapidly. In our model, the formation of molecular hydrogen is not efficient for metal-free gas, even for timescales larger than $1 \, \mathrm{Gyr}$. However, the efficiency increases rapidly as the ISM gets chemically enriched at levels of the order of $0.1 \, \mathrm{Z_\odot}$, following an increase in the amount of molecular hydrogen. This allows the formation of an extended gas disc of molecular material. 

Our model predicts a correlation between the molecular gas and SFR density, similar to the molecular KS law. It is worth noting that this relation is not imposed in our model but results naturally from the system of equations of our sub-grid treatment. The correlation between the $\Sigma_\mathrm{H_2}$ and $\Sigma_\mathrm{SFR}$ in Au6\_MOL is somewhat steeper than the observational estimates obtained from the PHANGS–ALMA sample \citep{Sun2023}.

An advantage of our model is that gas particles can have varying SFEs that depend on the local ISM properties rather than being an (imposed) fixed value. We find that the SFE per free fall time varies in the range $\sim \! 0.1\%$ to $\sim \! 10\%$ with typical values of $0.4-0.6\%$, except for very early times with a median of $0.08\%$ (measured at $t = 2 \, \mathrm{Gyr}$). The SFE per free-fall time is an increasing function of metallicity and density and shows a mild evolution for times $\gtrsim 3-4 \, \mathrm{Gyr}$. The obtained values for the SFEs are consistent with those derived from observations of molecular gas and from simulations of a turbulent ISM.  

We compared our model to the standard implementation linking SF to total gas density and to an alternative approach tying SF to $\mathrm{H}_2$ abundance, assuming a scaling with local gas pressure. The two simulations with $\mathrm{H}_2$-based SF laws show a delay in the SF onset of about $0.5 \, \mathrm{Gyr}$ with respect to the standard formulation. We find different distributions for the SFE per free-fall time of gas in the three runs, as well as different dependencies (if at all) on the properties of the ISM that might potentially be tested against observations.

In particular, our model does not impose any correlation between the SFR and the total or molecular gas surface densities and does not require an ad hoc efficiency parameter. In the current implementation, however, the molecular hydrogen is considered a precursor for SF, although this can be easily changed in the models where SF is linked to the atomic gas or a combination of the two components. Moreover, the dust model included here is extremely simple, and it will be extended in future work allowing for a better treatment of dust and its relation to molecular hydrogen and SF. Additional effects such as photodissociation of $\mathrm{H}_2$ by metagalactic Lyman-Werner radiation also need to be considered. Overall, our sub-grid model allows us to link the SF to the local ISM properties at the resolved scales, while being simple enough for simulating galaxy formation in a cosmological context, ignoring the high complexity of the ISM at $\lesssim 100 \, \mathrm{pc}$ scales.

\begin{acknowledgements}
    We thank the anonymous referee for their helpful comments and suggestions, which improved our paper. 
    EL acknowledges funding from the HORIZON-MSCA-2021-SE-01 Research and Innovation Programme under the Marie Skłodowska-Curie grant agreement No. 101086388 (LACEGAL-III: Latin American–Chinese–European Galaxy Formation Network). CS acknowledges stimulating and interesting discussions with Andi Burkert, Vadim Semenov, Andrey Kravtsov, and Oscar Agertz.  
    CS and SEN are members of the Carrera del Investigador Científico of CONICET. They acknowledge funding from Agencia Nacional de Promoción Científica y Tecnológica (PICT-201-0667).
\end{acknowledgements}

\bibliographystyle{aa}
\bibliography{bibliography}

\begin{thebibliography}{80}
\expandafter\ifx\csname natexlab\endcsname\relax\def\natexlab#1{#1}\fi

\bibitem[{Agertz {et~al.}(2013)Agertz, Kravtsov, Leitner, \& Gnedin}]{Agertz2013}
Agertz, O., Kravtsov, A.~V., Leitner, S.~N., \& Gnedin, N.~Y. 2013, The Astrophysical Journal, 770, 25

\bibitem[{Baker {et~al.}(2022)Baker, Maiolino, Belfiore, Bluck, Curti, Wylezalek, Bertemes, Bothwell, Lin, Thorp, \& Pan}]{Baker2022}
Baker, W.~M., Maiolino, R., Belfiore, F., {et~al.} 2022, Monthly Notices of the Royal Astronomical Society, 518, 4767

\bibitem[{Baker {et~al.}(2021)Baker, Maiolino, Bluck, Lin, Ellison, Belfiore, Pan, \& Thorp}]{Baker2021}
Baker, W.~M., Maiolino, R., Bluck, A. F.~L., {et~al.} 2021, Monthly Notices of the Royal Astronomical Society, 510, 3622

\bibitem[{Barnes \& Hut(1986)}]{Barnes1986}
Barnes, J. \& Hut, P. 1986, Nature, 324, 446

\bibitem[{Bigiel {et~al.}(2008)Bigiel, Leroy, Walter, Brinks, Blok, Madore, \& Thornley}]{Bigiel2008}
Bigiel, F., Leroy, A., Walter, F., {et~al.} 2008, The Astronomical Journal, 136, 2846

\bibitem[{Bigiel {et~al.}(2010)Bigiel, Walter, Blitz, Brinks, Blok, \& Madore}]{Bigiel2010}
Bigiel, F., Walter, F., Blitz, L., {et~al.} 2010, The Astronomical Journal, 140, 1194

\bibitem[{Black \& van Dishoeck(1987)}]{Black1987}
Black, J.~H. \& van Dishoeck, E.~F. 1987, The Astrophysical Journal, 322, 412

\bibitem[{Blitz \& Rosolowsky(2006)}]{Blitz2006}
Blitz, L. \& Rosolowsky, E. 2006, The Astrophysical Journal, 650, 933

\bibitem[{Blumenthal {et~al.}(1984)Blumenthal, Faber, Primack, \& Rees}]{Blumenthal1984}
Blumenthal, G.~R., Faber, S.~M., Primack, J.~R., \& Rees, M.~J. 1984, Nature, 311, 517

\bibitem[{Bolatto {et~al.}(2011)Bolatto, Leroy, Jameson, Ostriker, Gordon, Lawton, Stanimirović, Israel, Madden, Hony, Sandstrom, Bot, Rubio, Winkler, Roman-Duval, van Loon, Oliveira, \& Indebetouw}]{Bolatto2011}
Bolatto, A.~D., Leroy, A.~K., Jameson, K., {et~al.} 2011, The Astrophysical Journal, 741, 12

\bibitem[{Chabrier(2003)}]{Chabrier2003}
Chabrier, G. 2003, The Astrophysical Journal, 586, L133

\bibitem[{Christensen {et~al.}(2012)Christensen, Quinn, Governato, Stilp, Shen, \& Wadsley}]{Christensen2012}
Christensen, C., Quinn, T., Governato, F., {et~al.} 2012, Monthly Notices of the Royal Astronomical Society, 425, 3058

\bibitem[{Collaboration {et~al.}(2014)Collaboration, Ade, Aghanim, Armitage-Caplan, Arnaud, Ashdown, Atrio-Barandela, Aumont, Baccigalupi, Banday, Barreiro, Bartlett, Battaner, Benabed, Benoît, Benoit-Lévy, Bernard, Bersanelli, Bielewicz, Bobin, Bock, Bonaldi, Bond, Borrill, Bouchet, Bridges, Bucher, Burigana, Butler, Calabrese, Cappellini, Cardoso, Catalano, Challinor, Chamballu, Chary, Chen, Chiang, Chiang, Christensen, Church, Clements, Colombi, Colombo, Couchot, Coulais, Crill, Curto, Cuttaia, Danese, Davies, Davis, de~Bernardis, de~Rosa, de~Zotti, Delabrouille, Delouis, Désert, Dickinson, Diego, Dolag, Dole, Donzelli, Doré, Douspis, Dunkley, Dupac, Efstathiou, Elsner, Enßlin, Eriksen, Finelli, Forni, Frailis, Fraisse, Franceschi, Gaier, Galeotta, Galli, Ganga, Giard, Giardino, Giraud-Héraud, Gjerløw, González-Nuevo, Górski, Gratton, Gregorio, Gruppuso, Gudmundsson, Haissinski, Hamann, Hansen, Hanson, Harrison, Henrot-Versillé, Hernández-Monteagudo, Herranz, Hildebrandt, Hivon, Hobson, Holmes,
  Hornstrup, Hou, Hovest, Huffenberger, Jaffe, Jaffe, Jewell, Jones, Juvela, Keihänen, Keskitalo, Kisner, Kneissl, Knoche, Knox, Kunz, Kurki-Suonio, Lagache, Lähteenmäki, Lamarre, Lasenby, Lattanzi, Laureijs, Lawrence, Leach, Leahy, Leonardi, León-Tavares, Lesgourgues, Lewis, Liguori, Lilje, Linden-Vørnle, López-Caniego, Lubin, Macías-Pérez, Maffei, Maino, Mandolesi, Maris, Marshall, Martin, Martínez-González, Masi, Massardi, Matarrese, Matthai, Mazzotta, Meinhold, Melchiorri, Melin, Mendes, Menegoni, Mennella, Migliaccio, Millea, Mitra, Miville-Deschênes, Moneti, Montier, Morgante, Mortlock, Moss, Munshi, Murphy, Naselsky, Nati, Natoli, Netterfield, Nørgaard-Nielsen, Noviello, Novikov, Novikov, O’Dwyer, Osborne, Oxborrow, Paci, Pagano, Pajot, Paladini, Paoletti, Partridge, Pasian, Patanchon, Pearson, Pearson, Peiris, Perdereau, Perotto, Perrotta, Pettorino, Piacentini, Piat, Pierpaoli, Pietrobon, Plaszczynski, Platania, Pointecouteau, Polenta, Ponthieu, Popa, Poutanen, Pratt, Prézeau, Prunet,
  Puget, Rachen, Reach, Rebolo, Reinecke, Remazeilles, Renault, Ricciardi, Riller, Ristorcelli, Rocha, Rosset, Roudier, Rowan-Robinson, Rubiño-Martín, Rusholme, Sandri, Santos, Savelainen, Savini, Scott, Seiffert, Shellard, Spencer, Starck, Stolyarov, Stompor, Sudiwala, Sunyaev, Sureau, Sutton, Suur-Uski, Sygnet, Tauber, Tavagnacco, Terenzi, Toffolatti, Tomasi, Tristram, Tucci, Tuovinen, Türler, Umana, Valenziano, Valiviita, Van~Tent, Vielva, Villa, Vittorio, Wade, Wandelt, Wehus, White, White, Wilkinson, Yvon, Zacchei, \& Zonca}]{PlanckCollaboration2014}
Collaboration, P., Ade, P. A.~R., Aghanim, N., {et~al.} 2014, Astronomy and Astrophysics, 571, A16

\bibitem[{Dalgarno \& McCray(1972)}]{Dalgarno1972}
Dalgarno, A. \& McCray, R.~A. 1972, Annual Review of Astronomy and Astrophysics, 10, 375

\bibitem[{Draine \& Bertoldi(1996)}]{Draine1996}
Draine, B.~T. \& Bertoldi, F. 1996, The Astrophysical Journal, 468, 269

\bibitem[{Efstathiou(1992)}]{Efstathiou1992}
Efstathiou, G. 1992, Monthly Notices of the Royal Astronomical Society, 256, 43P

\bibitem[{Glover \& Clark(2012)}]{Glover2012}
Glover, S. C.~O. \& Clark, P.~C. 2012, Monthly Notices of the Royal Astronomical Society, 421, 9

\bibitem[{Glover \& Jappsen(2007)}]{Glover2007}
Glover, S. C.~O. \& Jappsen, A.~K. 2007, The Astrophysical Journal, 666, 1

\bibitem[{Gnedin {et~al.}(2009)Gnedin, Tassis, \& Kravtsov}]{Gnedin2009}
Gnedin, N.~Y., Tassis, K., \& Kravtsov, A.~V. 2009, The Astrophysical Journal, 697, 55

\bibitem[{Grand {et~al.}(2017)Grand, Gómez, Marinacci, Pakmor, Springel, Campbell, Frenk, Jenkins, \& White}]{Grand2017}
Grand, R. J.~J., Gómez, F.~A., Marinacci, F., {et~al.} 2017, Monthly Notices of the Royal Astronomical Society, 467, 179

\bibitem[{Hollenbach \& Salpeter(1971)}]{Hollenbach1971a}
Hollenbach, D. \& Salpeter, E.~E. 1971, The Astrophysical Journal, 163, 155

\bibitem[{Hollenbach {et~al.}(1971)Hollenbach, Werner, \& Salpeter}]{Hollenbach1971b}
Hollenbach, D.~J., Werner, M.~W., \& Salpeter, E.~E. 1971, The Astrophysical Journal, 163, 165

\bibitem[{Hopkins {et~al.}(2014)Hopkins, Kereš, Oñorbe, Faucher-Giguère, Quataert, Murray, \& Bullock}]{Hopkins2014}
Hopkins, P.~F., Kereš, D., Oñorbe, J., {et~al.} 2014, Monthly Notices of the Royal Astronomical Society, 445, 581

\bibitem[{Hultman \& Pharasyn(1999)}]{Hultman1999}
Hultman, J. \& Pharasyn, A. 1999, Astronomy and Astrophysics, 347, 769

\bibitem[{Iza {et~al.}(2022)Iza, Scannapieco, Nuza, Grand, Gómez, Springel, Pakmor, \& Marinacci}]{Iza2022}
Iza, F.~G., Scannapieco, C., Nuza, S.~E., {et~al.} 2022, Monthly Notices of the Royal Astronomical Society, 517, 832

\bibitem[{Katz(1992)}]{Katz1992}
Katz, N. 1992, The Astrophysical Journal, 391, 502

\bibitem[{Kennicutt(1998)}]{Kennicutt1998}
Kennicutt, R. C.~J. 1998, The Astrophysical Journal, 498, 541

\bibitem[{Kim {et~al.}(2013)Kim, Abel, Agertz, Bryan, Ceverino, Christensen, Conroy, Dekel, Gnedin, Goldbaum, Guedes, Hahn, Hobbs, Hopkins, Hummels, Iannuzzi, Keres, Klypin, Kravtsov, Krumholz, Kuhlen, Leitner, Madau, Mayer, Moody, Nagamine, Norman, Onorbe, O'Shea, Pillepich, Primack, Quinn, Read, Robertson, Rocha, Rudd, Shen, Smith, Szalay, Teyssier, Thompson, Todoroki, Turk, Wadsley, Wise, \& Zolotov}]{Kim2013}
Kim, J.-h., Abel, T., Agertz, O., {et~al.} 2013, The Astrophysical Journal Supplement Series, 210, 14

\bibitem[{Kim {et~al.}(2016)Kim, Agertz, Teyssier, Butler, Ceverino, Choi, Feldmann, Keller, Lupi, Quinn, Revaz, Wallace, Gnedin, Leitner, Shen, Smith, Thompson, Turk, Abel, Arraki, Benincasa, Chakrabarti, DeGraf, Dekel, Goldbaum, Hopkins, Hummels, Klypin, Li, Madau, Mandelker, Mayer, Nagamine, Nickerson, O’Shea, Primack, Roca-Fàbrega, Semenov, Shimizu, Simpson, Todoroki, Wadsley, \& Wise}]{Kim2016}
Kim, J.-h., Agertz, O., Teyssier, R., {et~al.} 2016, The Astrophysical Journal, 833, 202

\bibitem[{Krumholz \& Gnedin(2011)}]{Krumholz2011}
Krumholz, M.~R. \& Gnedin, N.~Y. 2011, The Astrophysical Journal, 729, 36

\bibitem[{Krumholz \& McKee(2005)}]{Krumholz2005}
Krumholz, M.~R. \& McKee, C.~F. 2005, The Astrophysical Journal, 630, 250

\bibitem[{Krumholz {et~al.}(2019)Krumholz, McKee, \& Bland-Hawthorn}]{Krumholz2019}
Krumholz, M.~R., McKee, C.~F., \& Bland-Hawthorn, J. 2019, Annual Review of Astronomy and Astrophysics, 57, 227

\bibitem[{Krumholz {et~al.}(2009)Krumholz, McKee, \& Tumlinson}]{Krumholz2009}
Krumholz, M.~R., McKee, C.~F., \& Tumlinson, J. 2009, The Astrophysical Journal, 699, 850

\bibitem[{Leroy {et~al.}(2008)Leroy, Walter, Brinks, Bigiel, Blok, Madore, \& Thornley}]{Leroy2008}
Leroy, A.~K., Walter, F., Brinks, E., {et~al.} 2008, The Astronomical Journal, 136, 2782

\bibitem[{Leroy {et~al.}(2013)Leroy, Walter, Sandstrom, Schruba, Munoz-Mateos, Bigiel, Bolatto, Brinks, de~Blok, Meidt, Rix, Rosolowsky, Schinnerer, Schuster, \& Usero}]{Leroy2013}
Leroy, A.~K., Walter, F., Sandstrom, K., {et~al.} 2013, The Astronomical Journal, 146, 19

\bibitem[{Libeskind {et~al.}(2020)Libeskind, Carlesi, Grand, Khalatyan, Knebe, Pakmor, Pilipenko, Pawlowski, Sparre, Tempel, Wang, Courtois, Gottlöber, Hoffman, Minchev, Pfrommer, Sorce, Springel, Steinmetz, Tully, Vogelsberger, \& Yepes}]{Libeskind2020}
Libeskind, N.~I., Carlesi, E., Grand, R. J.~J., {et~al.} 2020, Monthly Notices of the Royal Astronomical Society, 498, 2968

\bibitem[{Marinacci {et~al.}(2013)Marinacci, Pakmor, \& Springel}]{Marinacci2013}
Marinacci, F., Pakmor, R., \& Springel, V. 2013, Monthly Notices of the Royal Astronomical Society, 437, 1750

\bibitem[{Matteucci(2003)}]{Matteucci2003}
Matteucci, F. 2003, The Chemical Evolution of the Galaxy (Springer Dordrecht)

\bibitem[{McKee \& Ostriker(1977)}]{McKee1977}
McKee, C.~F. \& Ostriker, J.~P. 1977, The Astrophysical Journal, 218, 148

\bibitem[{Millán-Irigoyen {et~al.}(2020)Millán-Irigoyen, Mollá, \& Ascasibar}]{Millan-Irigoyen2020}
Millán-Irigoyen, I., Mollá, M., \& Ascasibar, Y. 2020, Monthly Notices of the Royal Astronomical Society, 494, 146

\bibitem[{Mollá {et~al.}(2015)Mollá, Cavichia, Gavilán, \& Gibson}]{Molla2015}
Mollá, M., Cavichia, O., Gavilán, M., \& Gibson, B.~K. 2015, Monthly Notices of the Royal Astronomical Society, 451, 3693

\bibitem[{Mollá {et~al.}(2017)Mollá, Díaz, Ascasibar, \& Gibson}]{Molla2017}
Mollá, M., Díaz, A.~I., Ascasibar, Y., \& Gibson, B.~K. 2017, Monthly Notices of the Royal Astronomical Society, 468, 305

\bibitem[{Mollá {et~al.}(2009)Mollá, García-Vargas, \& Bressan}]{Molla2009}
Mollá, M., García-Vargas, M.~L., \& Bressan, A. 2009, Monthly Notices of the Royal Astronomical Society, 398, 451

\bibitem[{Murante {et~al.}(2014)Murante, Monaco, Borgani, Tornatore, Dolag, \& Goz}]{Murante2014}
Murante, G., Monaco, P., Borgani, S., {et~al.} 2014, Monthly Notices of the Royal Astronomical Society, 447, 178

\bibitem[{Murante {et~al.}(2010)Murante, Monaco, Giovalli, Borgani, \& Diaferio}]{Murante2010}
Murante, G., Monaco, P., Giovalli, M., Borgani, S., \& Diaferio, A. 2010, Monthly Notices of the Royal Astronomical Society, 405, 1491

\bibitem[{Naab \& Ostriker(2017)}]{Naab2017}
Naab, T. \& Ostriker, J.~P. 2017, Annual Review of Astronomy and Astrophysics, 55, 59

\bibitem[{Nebrin(2023)}]{Nebrin2023}
Nebrin, O. 2023, Research Notes of the AAS, 7, 90

\bibitem[{Osterbrock \& Ferland(2006)}]{Osterbrock2006}
Osterbrock, D. \& Ferland, G. 2006, Astrophysics Of Gas Nebulae and Active Galactic Nuclei (University Science Books)

\bibitem[{Pagel(2009)}]{Pagel2009}
Pagel, B. E.~J. 2009, Nucleosynthesis and Chemical Evolution of Galaxies, 2nd edn. (Cambridge University Press)

\bibitem[{Pakmor {et~al.}(2018)Pakmor, Guillet, Pfrommer, Gómez, Grand, Marinacci, Simpson, \& Springel}]{Pakmor2018}
Pakmor, R., Guillet, T., Pfrommer, C., {et~al.} 2018, Monthly Notices of the Royal Astronomical Society, 481, 4410

\bibitem[{Pakmor {et~al.}(2017)Pakmor, Gómez, Grand, Marinacci, Simpson, Springel, Campbell, Frenk, Guillet, Pfrommer, \& White}]{Pakmor2017}
Pakmor, R., Gómez, F.~A., Grand, R. J.~J., {et~al.} 2017, Monthly Notices of the Royal Astronomical Society, 469, 3185

\bibitem[{Pakmor {et~al.}(2014)Pakmor, Marinacci, \& Springel}]{Pakmor2014}
Pakmor, R., Marinacci, F., \& Springel, V. 2014, The Astrophysical Journal Letters, 783, L20

\bibitem[{Pakmor {et~al.}(2015)Pakmor, Springel, Bauer, Mocz, Munoz, Ohlmann, Schaal, \& Zhu}]{Pakmor2015}
Pakmor, R., Springel, V., Bauer, A., {et~al.} 2015, Monthly Notices of the Royal Astronomical Society, 455, 1134

\bibitem[{Pelupessy {et~al.}(2006)Pelupessy, Papadopoulos, \& van~der Werf}]{Pelupessy2006}
Pelupessy, F.~I., Papadopoulos, P.~P., \& van~der Werf, P. 2006, The Astrophysical Journal, 645, 1024

\bibitem[{Portinari {et~al.}(1998)Portinari, Chiosi, \& Bressan}]{Portinari1998}
Portinari, L., Chiosi, C., \& Bressan, A. 1998, Astronomy and Astrophysics, 334, 505

\bibitem[{Rees \& Ostriker(1977)}]{Rees1977}
Rees, M.~J. \& Ostriker, J.~P. 1977, Monthly Notices of the Royal Astronomical Society, 179, 541

\bibitem[{Robertson \& Kravtsov(2008)}]{Robertson2008}
Robertson, B.~E. \& Kravtsov, A.~V. 2008, The Astrophysical Journal, 680, 1083

\bibitem[{Roychowdhury {et~al.}(2015)Roychowdhury, Huang, Kauffmann, Wang, \& Chengalur}]{Roychowdhury2015}
Roychowdhury, S., Huang, M.-L., Kauffmann, G., Wang, J., \& Chengalur, J.~N. 2015, Monthly Notices of the Royal Astronomical Society, 449, 3700

\bibitem[{Scannapieco {et~al.}(2006)Scannapieco, Tissera, White, \& Springel}]{Scannapieco2006}
Scannapieco, C., Tissera, P.~B., White, S. D.~M., \& Springel, V. 2006, Monthly Notices of the Royal Astronomical Society, 371, 1125

\bibitem[{Scannapieco {et~al.}(2012)Scannapieco, Wadepuhl, Parry, Navarro, Jenkins, Springel, Teyssier, Carlson, Couchman, Crain, Vecchia, Frenk, Kobayashi, Monaco, Murante, Okamoto, Quinn, Schaye, Stinson, Theuns, Wadsley, White, \& Woods}]{Scannapieco2012}
Scannapieco, C., Wadepuhl, M., Parry, O.~H., {et~al.} 2012, Monthly Notices of the Royal Astronomical Society, 423, 1726

\bibitem[{Schinnerer \& Leroy(2024)}]{Schinnerer2024}
Schinnerer, E. \& Leroy, A.~K. 2024, Annual Review of Astronomy and Astrophysics, 62, 369

\bibitem[{Schmidt(1959)}]{Schmidt1959}
Schmidt, M. 1959, The Astrophysical Journal, 129, 243

\bibitem[{Schmidt(1963)}]{Schmidt1963}
Schmidt, M. 1963, The Astrophysical Journal, 137, 758

\bibitem[{Schruba {et~al.}(2011)Schruba, Leroy, Walter, Bigiel, Brinks, de~Blok, Dumas, Kramer, Rosolowsky, Sandstrom, Schuster, Usero, Weiss, \& Wiesemeyer}]{Schruba2011}
Schruba, A., Leroy, A.~K., Walter, F., {et~al.} 2011, The Astronomical Journal, 142, 37

\bibitem[{Semenov {et~al.}(2016)Semenov, Kravtsov, \& Gnedin}]{Semenov2016}
Semenov, V.~A., Kravtsov, A.~V., \& Gnedin, N.~Y. 2016, The Astrophysical Journal, 826, 200

\bibitem[{Silk(1977)}]{Silk1977}
Silk, J. 1977, The Astrophysical Journal, 211, 638

\bibitem[{Springel(2005)}]{Springel2005}
Springel, V. 2005, Monthly Notices of the Royal Astronomical Society, 364, 1105

\bibitem[{Springel(2010)}]{Springel2010}
Springel, V. 2010, Monthly Notices of the Royal Astronomical Society, 401, 791

\bibitem[{Springel \& Hernquist(2003)}]{Springel2003}
Springel, V. \& Hernquist, L. 2003, Monthly Notices of the Royal Astronomical Society, 339, 289

\bibitem[{Springel {et~al.}(2018)Springel, Pakmor, Pillepich, Weinberger, Nelson, Hernquist, Vogelsberger, Genel, Torrey, Marinacci, \& Naiman}]{Springel2018}
Springel, V., Pakmor, R., Pillepich, A., {et~al.} 2018, Monthly Notices of the Royal Astronomical Society, 475, 676

\bibitem[{Steinmetz \& Mueller(1994)}]{Steinmetz1994}
Steinmetz, M. \& Mueller, E. 1994, Astronomy and Astrophysics, 281, L97

\bibitem[{Sun {et~al.}(2023)Sun, Leroy, Ostriker, Meidt, Rosolowsky, Schinnerer, Wilson, Utomo, Belfiore, Blanc, Emsellem, Faesi, Groves, Hughes, Koch, Kreckel, Liu, Pan, Pety, Querejeta, Razza, Saito, Sardone, Usero, Williams, Bigiel, Bolatto, Chevance, Dale, Gensior, Glover, Grasha, Henshaw, Jiménez-Donaire, Klessen, Kruijssen, Murphy, Neumann, Teng, \& Thilker}]{Sun2023}
Sun, J., Leroy, A.~K., Ostriker, E.~C., {et~al.} 2023, The Astrophysical Journal Letters, 945, L19

\bibitem[{Utomo {et~al.}(2018)Utomo, Sun, Leroy, Kruijssen, Schinnerer, Schruba, Bigiel, Blanc, Chevance, Emsellem, Herrera, Hygate, Kreckel, Ostriker, Pety, Querejeta, Rosolowsky, Sandstrom, \& Usero}]{Utomo2018}
Utomo, D., Sun, J., Leroy, A.~K., {et~al.} 2018, The Astrophysical Journal Letters, 861, L18

\bibitem[{Valentini {et~al.}(2022)Valentini, Dolag, Borgani, Murante, Maio, Tornatore, Granato, Ragone-Figueroa, Burkert, Ragagnin, \& Rasia}]{Valentini2022}
Valentini, M., Dolag, K., Borgani, S., {et~al.} 2022, Monthly Notices of the Royal Astronomical Society, 518, 1128

\bibitem[{Vogelsberger {et~al.}(2013)Vogelsberger, Genel, Sijacki, Torrey, Springel, \& Hernquist}]{Vogelsberger2013}
Vogelsberger, M., Genel, S., Sijacki, D., {et~al.} 2013, Monthly Notices of the Royal Astronomical Society, 436, 3031

\bibitem[{White \& Frenk(1991)}]{White1991}
White, S. D.~M. \& Frenk, C.~S. 1991, The Astrophysical Journal, 379, 52

\bibitem[{White \& Rees(1978)}]{White1978}
White, S. D.~M. \& Rees, M.~J. 1978, Monthly Notices of the Royal Astronomical Society, 183, 341

\bibitem[{Wolfire {et~al.}(2008)Wolfire, Tielens, Hollenbach, \& Kaufman}]{Wolfire2008}
Wolfire, M.~G., Tielens, A. G. G.~M., Hollenbach, D., \& Kaufman, M.~J. 2008, The Astrophysical Journal, 680, 384

\bibitem[{Wong \& Blitz(2002)}]{Wong2002}
Wong, T. \& Blitz, L. 2002, The Astrophysical Journal, 569, 157

\bibitem[{Yepes {et~al.}(1997)Yepes, Kates, Khokhlov, \& Klypin}]{Yepes1997}
Yepes, G., Kates, R., Khokhlov, A., \& Klypin, A. 1997, Monthly Notices of the Royal Astronomical Society, 284, 235

\end{thebibliography}

\begin{appendix}

\section{Predictions of the model for fixed initial conditions}
\label{app:fixed_ics}

In this section, we describe the behaviour of Eqs.\ (\ref{eq:odes}) for fixed ICs, which cover the range of parameters (density, metallicity, and relative fraction of ionised and atomic gas) typical of our galaxy formation simulations and the adopted resolution (which sets the maximum gas densities achieved). A first approach to understanding the behaviour of the model equations which govern the evolution of the gas and stellar phases was given in Section~\ref{subsec:implementation}, where we compared the typical timescales involved in the transformation of mass between the different phases. As explained in the text, the three timescales of our model -- i.e. recombination, condensation, and SF -- have a similar dependency on the density ($\tau \propto \rho^{-1}$ for recombination and condensation, and $\tau_\mathrm{star} \propto \rho^{-1/2}$ for SF); the recombination timescale also depends on the ionised fraction, and the condensation timescale has additional dependencies on the gas metallicity and the stellar fraction (which sets the value of $f_i + f_a + f_m$, see Eq. (\ref{eq:tau_cond})). Furthermore, $\eta_\mathrm{diss}$, $\eta_\mathrm{ion}$, and $R$ depend on the gas metallicity.

\begin{figure*}[ht]
    \centering
    \includegraphics[width=17cm]{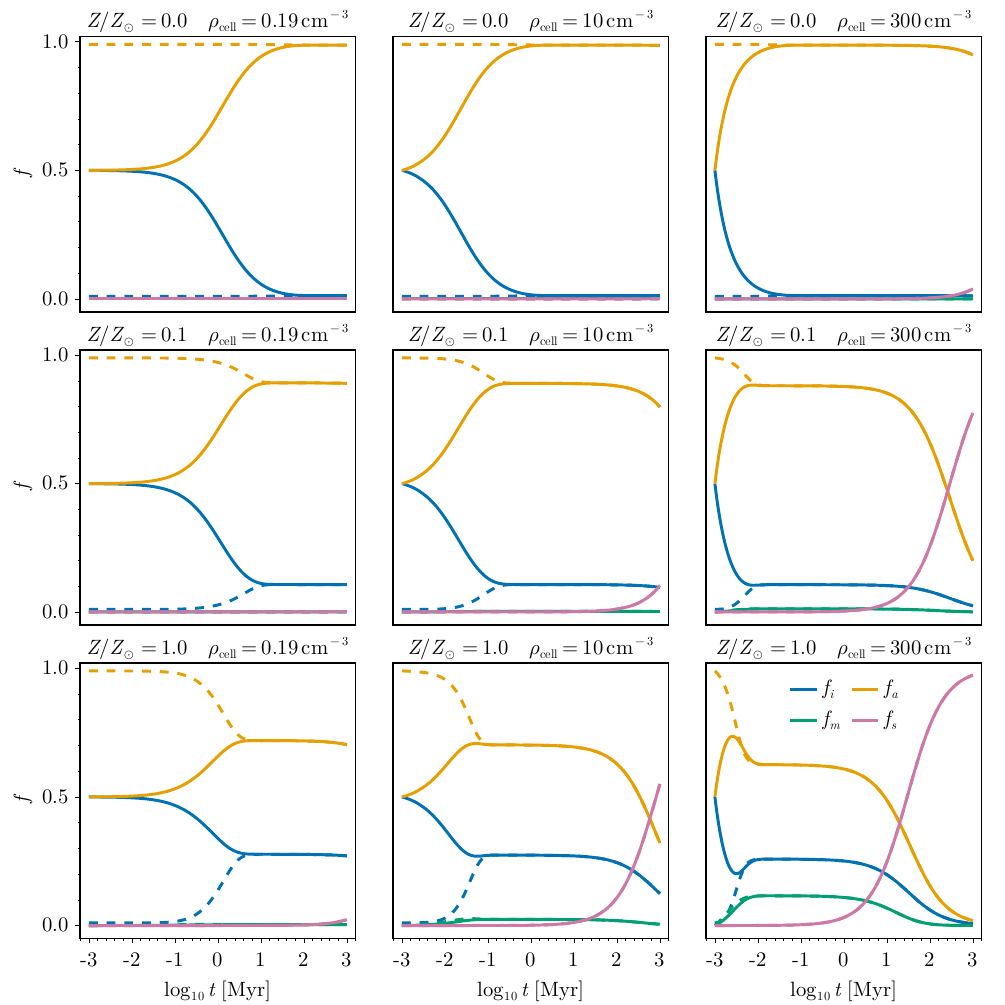}
    \caption{Evolution of the gas phases when integrating Eqs.\ (\ref{eq:odes}) up to $t = 1 \, \mathrm{Gyr}$. Different panels corresponds to different values of gas metallicity and density, as indicated above each panel. Solid lines correspond to the ICs $f_i = f_a = 0.5$, and dashed lines with $f_i = 0.01$ and $f_a = 0.99$. In both cases the initial values for $f_m$ and $f_s$ are $0$.}
    \label{fig:FracvsTime}
\end{figure*}

\begin{figure*}[ht]
    \centering
    \includegraphics[width=17cm]{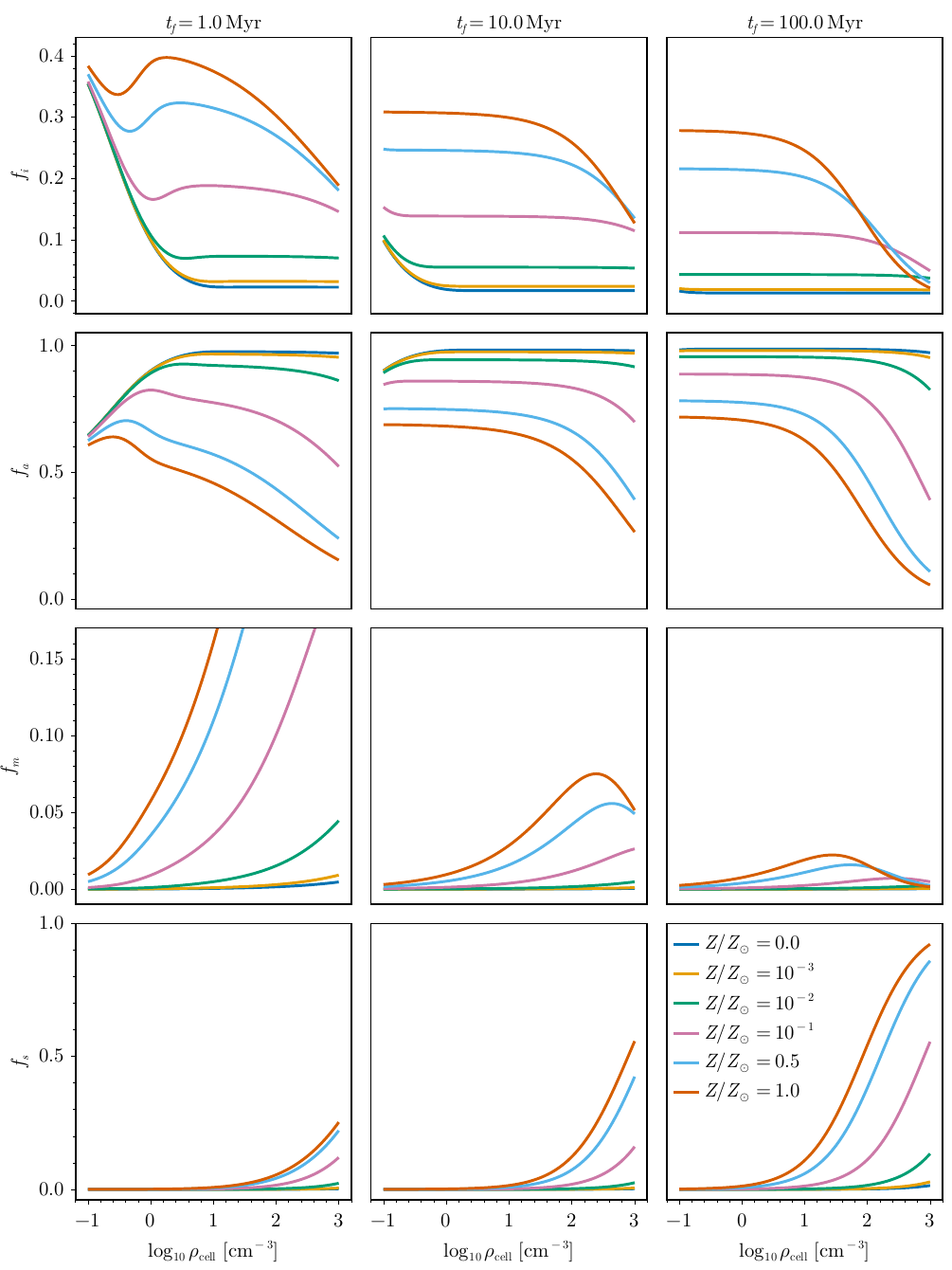}
    \caption{Final value of the different phases (rows) when integrating Eqs.\ (\ref{eq:odes}) for $1$, $10$, and $100 \, \mathrm{Myr}$ (columns). Each colour corresponds to a different value of metallicity.}
    \label{fig:fractions_vs_cell-density}
\end{figure*}

It is important to note that, in a simulation, as we integrate our system of equations in time, the timescales vary. In the results shown in this section, the density and metallicity are kept fixed during integration, unlike $\tau_\mathrm{rec}$ and $\tau_\mathrm{cond}$, which depend on the variables, ionised and stellar fractions, respectively. In Fig.~\ref{fig:FracvsTime} we show the time-evolution of the various fractions, for different initial metallicities and densities (see Fig.~\ref{fig:fractions_vs_cell-density} for understanding the selected density/metallicity values shown here). We show results up to $1 \, \mathrm{Gyr}$, to understand the predictions of the model up to times where star particles start to form at various physical conditions. It is worth noting that in our cosmological simulation, all the gas cells that have been transformed into stars were integrated for a total time $< 10 \, \mathrm{Myr}$.  

Let us first focus on the solid lines, which show a case where the initial atomic and ionised fractions are equal (with the stellar and molecular phases set to zero). The Fig.~\ref{fig:FracvsTime} shows a first transition, from ionised to atomic material, at times determined by the recombination timescale $\tau_\mathrm{rec}$ (see Fig.~\ref{fig:timescale_comparison}). This timescale is inversely proportional to the density, explaining why the transition occurs earlier in the right-hand panels (i.e. the highest density ones) and later for the left-hand ones. A second transition occurs from the atomic to the molecular phase, regulated by $\tau_\mathrm{cond}$. In this case, the transition occurs earlier as we go to higher densities, where the transition also becomes more efficient. This results in a significant production of stellar mass. For the metallicity dependence, as the gas gets enriched (from the top to the bottom rows) only $\tau_\mathrm{cond}$ is affected, such that higher molecular/stellar fractions are obtained. Once a stellar phase is formed, the additional dependencies regulating the processes of photodissociation, photoionisation, and mass recycling will make our system of equations even more complex.

Finally, the dashed lines in this figure show what happens if we start from a different IC where almost all material is in the atomic phase ($f_a(0) = 0.99$ and $f_i(0) = 0.01$). In this case, we see that results converge quite fast to the same values as the case where the initial atomic and ionised phases are $0.5$. This means that our system of equations has low sensitivity to the ICs (not so much with the parameters $\rho_\mathrm{cell}$ and $Z$), and in all cases the stellar fraction starts growing only after this convergence takes place.

In Fig.~\ref{fig:FracvsTime} we showed the predictions for our model for a set of fixed ICs, and discussed that the initial ionised/atomic fractions are unimportant considering the relevant timescales of the model. In Fig.~\ref{fig:fractions_vs_cell-density} we better show how predictions depend on the gas density and metallicity. The range of densities has been selected between the density threshold for SF (which determines the minimum density a gas cells must have to form stars, and thus to enter our routine) and the typical maximum density seen in our simulation (which depend on numerical resolution). In the case of metallicity, we cover from metal-free to highly enriched gas. We present in Fig.~\ref{fig:fractions_vs_cell-density} the fractions of ionised, atomic, molecular, and stellar phases, as a function of density and for various metallicities (note the different y-ranges of the various rows, which were chosen for the sake of clarity). The three columns correspond to different integration times of $1$, $10$, and $100 \, \mathrm{Myr}$.

The behaviour of the different phases as a function of density is determined by the dependence of all timescales on density: the larger the density, the smaller the timescales; depending on the time observed, different transitions have already taken place. In general, the ionised and atomic fractions are more efficiently transformed into the molecular/stellar phases for higher densities, and this effect is enhanced as the system evolves for longer times. The dependence on metallicity is such that metal-free gas is unable to produce a significant amount of stellar mass; however, as the metallicity reaches about $10^{-1} \, Z_\odot$, the fraction of molecular gas increases significantly, and so does the fraction of stellar mass (which in turn, consumes the molecular gas).

The results of this appendix show that, depending on the time-evolution of our equations, equilibrium solutions can be valid for a given period of time; however, as the ICs of each gas cell vary with time, as well as the timescales involved in the equations (which depend on the properties at the time a particle enters the routine), it is not trivial to predict the impact of our SF prescription in a cosmological simulation.

\end{appendix}

\end{document}